\documentclass[]{raa}            
\usepackage{graphicx,times}  
\pdfoutput=1           
\usepackage{natbib}
\usepackage{amssymb,amsmath}
\usepackage{longtable}

\bibpunct{(}{)}{;}{a}{}{,}
\usepackage{ulem}

\usepackage[a4paper=true,pagebackref=true]{hyperref}
\begin{document}

   \title{Investigating Variable Stars in the Open Cluster NGC 1912 and Its Surrounding Field
}

 \volnopage{ {\bf 20XX} Vol.\ {\bf X} No. {\bf XX}, 000--000}
   \setcounter{page}{1}

  \author{Chunyan Li\inst{1, 2}, Ali Esamdin\inst{1}, Yu Zhang\inst{1}, Fangfang Song\inst{1, 2}, Xiangyun Zeng\inst{1, 2}, Li Chen\inst{2,3}, Hubiao Niu\inst{1,4}, Jianying Bai\inst{1,2}, Junhui Liu\inst{1}
}
\institute{Xinjiang Astronomical Observatory, the Chinese Academy of Sciences, Urumqi 830011, China; {\it aliyi@xao.ac.cn}\\
\and University of Chinese Academy of Sciences, Beijing 100049, Beijing, China\\
\and Shanghai Astronomical Observatory, the Chinese Academy of Sciences, Shanghai 200030, China\\
\and Department of Astronomy, Beijing Normal University, Beijing 100875, China\\
}

\abstract{In this work, we studied the variable stars in the open cluster NGC 1912 based on the photometric observations and $Gaia$ DR2 data.
More than 3600 CCD frames in $B, V, R$ filters were reduced, and we obtained the light curves that span about 63 hours.
By analyzing these light curves, we detected 24 variable stars, including 16 periodic variable stars, seven eclipsing binaries and one star whose type is unclear.
Among these 24 variable stars, 11 are newly-discovered, which are classified as six $\gamma$\,Doradus stars, one $\delta$\,Scuti star, three detached binaries and one contact binaries.
We also confirmed 13 previously known variable stars.
Based on cluster members identified by Cantat-Gaudin et al. (2018), we inferred cluster memberships for these detected variable stars.
Using $Gaia$~DR2 data, we plotted a new color-magnitude diagram for NGC 1912, and showed the nature of variable cluster members in kinematical properties and heliocentric distance. 
Among the 24 variable stars, seven variables are probable cluster members, which show homogeneity in kinematic characters and space position with the established cluster members.
Four of the seven variable cluster members are the previously discovered stars, consisting of two $\gamma$\,Dor stars and two $\delta$\,Sct stars.
The remaining three variable cluster members, which are all $\gamma$\,Dor stars, are firstly detected in this work.
The main physical parameters of these variable cluster members estimated from the color-magnitude diagram are $log(age/yr)=8.75$, $[Fe/H]=-0.1$, $m-M=10.03$ mag, and E($B$-$V$) = 0.307.
\keywords{open clusters and associations: NGC 1912 -- techniques: photometric -- stars: variables: general}
}

   \authorrunning{C.-Y. Li et al. }            
   \titlerunning{Variable Stars in NGC 1912}
   \maketitle

%
\section{Introduction}\label{sec:Int}
Open clusters are ideal tracers to study the stellar populations in our Galaxy.
Stars in one open cluster originated in the same interstellar cloud, so they can be assumed to have similar heliocentric distance, age, and chemical composition \citep{Friel+1995, Piskunov+etal+2006}.
Searching for variable stars in the open cluster can not only verify the stellar evolution theory but also offers important clues for further understanding of the structure and the evolution of the Milky Way \citep{Piskunov+etal+2006}.
However, the critical matter is to identify the membership of detected variable stars.
The recent second $Gaia$ data release ($Gaia$ DR2) \citep{GaiaCollaboration+etal+2018} provides a catalog of 1.3 billion sources with high precision parameter measurements to study the probable cluster memberships of stars.

The open cluster NGC 1912 [RA(J2000.0)=$05^{h}28^{m}40^{s}$, Dec(J2000.0)=$30^{\circ}50^{'}54^{''}$] is situated in the anti-center direction of the Galaxy, in Auriga \citep{Pandey+etal+2007}.
Based on photometry and spectrometry observations over the last six decades, physical properties of NGC 1912 are reliably established with its distance, reddening and age to be $1400\pm100$ pc, $0.25\pm0.02$ mag and $300\pm80$ Myr \citep{Hoag+Applequist+1965, Subramaniam+Sagar+1999, Kharchenko+etal+2005, Pandey+etal+2007}, respectively.
\cite{Dias+etal+2002b} provided a new catalog of NGC 1912 which updated the previous catalogs of \cite{Lynga+1985} and listed the membership probability of candidates based on Tycho2 proper motions \citep{Dias+etal+2002a}.
 Recently, \cite{Cloutier+etal+2018} identified 807 members of NGC 1912 using high-precision astrometric data from $Gaia$ DR2.
 These members are located in an area with center ($\alpha_{J2000}=5^{h}28^{m}41^{s}$, $\delta_{J2000}=35^{\circ}51^{'}19^{''}$) and radius $r=30^{'}$.
\citet{Monteiro+Dias+2019} estimated the distance, reddening, and age of the cluster to be $990\pm24$ pc, $0.307\pm0.020$ mag, and $log(age/yr)=8.538\pm0.0034$ from the isochrone fitting of NGC 1912 members using {\it Gaia} DR2 data as well.

Different types of variable stars were detected in the open cluster NGC 1912 and its surrounding field, such as Delta Scuti ($\delta$\,Sct), Gamma Doradus ($\gamma$\,Dor), and eclipsing binaries (EBs).
$\delta$\,Sct stars are A- and F-type stars pulsating in $p$-modes with typical frequencies in the range $5-50\,\mathrm{d^{-1}}$ \citep{2000ASPC..210....3B}.
The $p$-modes in $\delta$\,Sct stars are excited by the opacity mechanism operating in the He II ionization zone, as well as by turbulent pressure \citep{Antoci2014T, Xiong2016T}.
$\gamma$\,Dor stars are early F-type  pulsating stars with multiple periods typically between 0.4 and 3 days \citep{1999PASP..111..840K}.
$\gamma$\,Dor stars exhibit $g$-modes pulsations which are excited by a convective flux modulation mechanism \citep{Guzik+etal+2000, Dupret+etal+2005b, Grigahc_ne_2010}.
EBs can be divided into Algol (EA), $\beta$ Lyrae (EB) and W UMa (EW) types based on the shape of their light curves \citep{Gerry2003O}.

\citet{Szabo+etal+2006} detected 14 variable stars in the field of NGC 1912.
There are one EW type variables, three EA type variables, five $\delta$\,Sct pulsators, four long-term variable stars, and one unknown type variable star.
$BV$ photometry of the cluster were performed by \cite{Jeon+2009} in a larger field of view around the center of NGC 1912.
They found 15 $\delta$\,Sct stars and two $\gamma$\,Dor stars, and confirmed three $\delta$\,Sct pulsators which were found by \citet{Szabo+etal+2006} .

Among the 20 detected variable stars, 14 variables were located within radius $30^{'}$ from the center of NGC 1912. 
Nevertheless, the memberships of these variable stars were not provided in the two above works.
High precision parameters of these variable stars are needed to confirm their memberships for NGC 1912.
For the study of variable stars, the more recently Transiting Exoplanet Survey Satellite (TESS) mission data \citep{2015JATIS...1a4003R} provide new insight into pulsation stars and eclipsing binaries \citep[e.g.][]{2019MNRASA, 2020MNRASG}.

In this paper, we present CCD time-series photometry around open cluster NGC 1912 in $B, V$ and $R$ bands to investigate and classify variable stars.
We also infer the probable cluster memberships of detected variable stars based on {\it Gaia} DR2.
In Section \ref{sec:O&A}, we describe the target observations and data reduction.
Methods we used to detect and classify variable stars and membership probabilities of the detected variable stars provided by \citet{Cantat-Gaudin+etal+2018} are presented in Section \ref{sec: results}.
In Section \ref{sec:discu}, we compared the  results of this work with the previous works, showed the properties of the variable cluster members, and discussed the possibility of a variable star being another cluster member.
We present a summary of results in Section \ref{sec: SM}.

\section{Observations and data reduction}\label{sec:O&A}

 We observed the cluster NGC 1912 during nine nights from December 1st to 16th in 2015 at the Nanshan station of the Xinjiang Astronomical Observatory.
The Nanshan 1-m telescope (NOWT), equipped with a E2V CCD203-82 ($4196 \times 4096$ pixels, pixel size of 12 $\mu$m), was used in our observations.
The angular resolution of the telescope was 1.125$^{''}$/pixel, yielding a filed of view of $1.3 ^{\circ}\times 1.3 ^{\circ}$.
An area of $2600 \times 2400$ or $2000 \times 2000$ pixels near the center of the CCD chip was read out,, corresponding to a $48.75^{'} \times 45^{'}$ or $37.5^{'} \times 37.5^{'}$ filed of view around the center of the cluster.
The CCD operates at about $-120^{\circ}$C with liquid nitrogen cooling thus the dark current is less than $1e^{-}pix^{-1}h^{-1}$ at $-120^{\circ}$C.
The time-series observations were made using Johnson-Cousin-Bessel $B, V$ and $R$ filters (12s, 9s and 8s exposures respectively).
In total, 1228, 1220 and 1227 CCD frames in $B, V$ and $R$ bands were obtained, respectively.
The full observing journal is listed in Table~\ref{tab:log}.

\begin{table*}
\begin{center}	
	\caption{Log of observations run in $B, V$ and $R$ bands.\label{tab:log}}
	\begin{tabular}{cccccc}
		\hline
		\hline
		Date & CCD & FOV & length & Frames & Exposure Time\\
		& & (arcmin$^{2}$) & (h) & $B, V, R$ & $B, V, R$\\
		\hline
2015 Nov 01 & $2600 \times 2400$ & $48.75 \times 45$ & 9 & 136,128,138 &12,9,8 \\
2015 Nov 02 & $2600 \times 2400$ & $48.75 \times 45$ &9 & 156,156,154 &12,9,8 \\
2015 Nov 03 & $2600 \times 2400$ & $48.75 \times 45$ & 9 &202,201,202 &12,9,8 \\
2015 Nov 04 & $2600 \times 2400$ & $48.75 \times 45$ & 5 & 110,111,110 &12,9,8 \\
2015 Nov 05 & $2600 \times 2400$ & $48.75 \times 45$ & 9 & 195,195,194 & 12,9,8 \\
2015 Nov 07 & $2600 \times 2400$ & $48.75 \times 45$ & 7 & 153,153,153 &12,9,8 \\
2015 Nov 09 & $2600 \times 2400$ & $48.75 \times 45$ & 7 & 144,146,144 & 12,9,8 \\
2015 Nov 15 & $2000 \times 2000$ & $37.5 \times 37.5$ & 5 & 78,80,79 &12,9,8 \\
2015 Nov 16 & $2000 \times 2000$ & $37.5 \times 37.5$ & 3 & 54,54,53 &12,9,8 \\
		\hline
	\end{tabular}
\end{center}
\end{table*}

\begin{figure}
   \begin{center}
	\includegraphics[height=8cm,width=9cm,clip,angle=0]{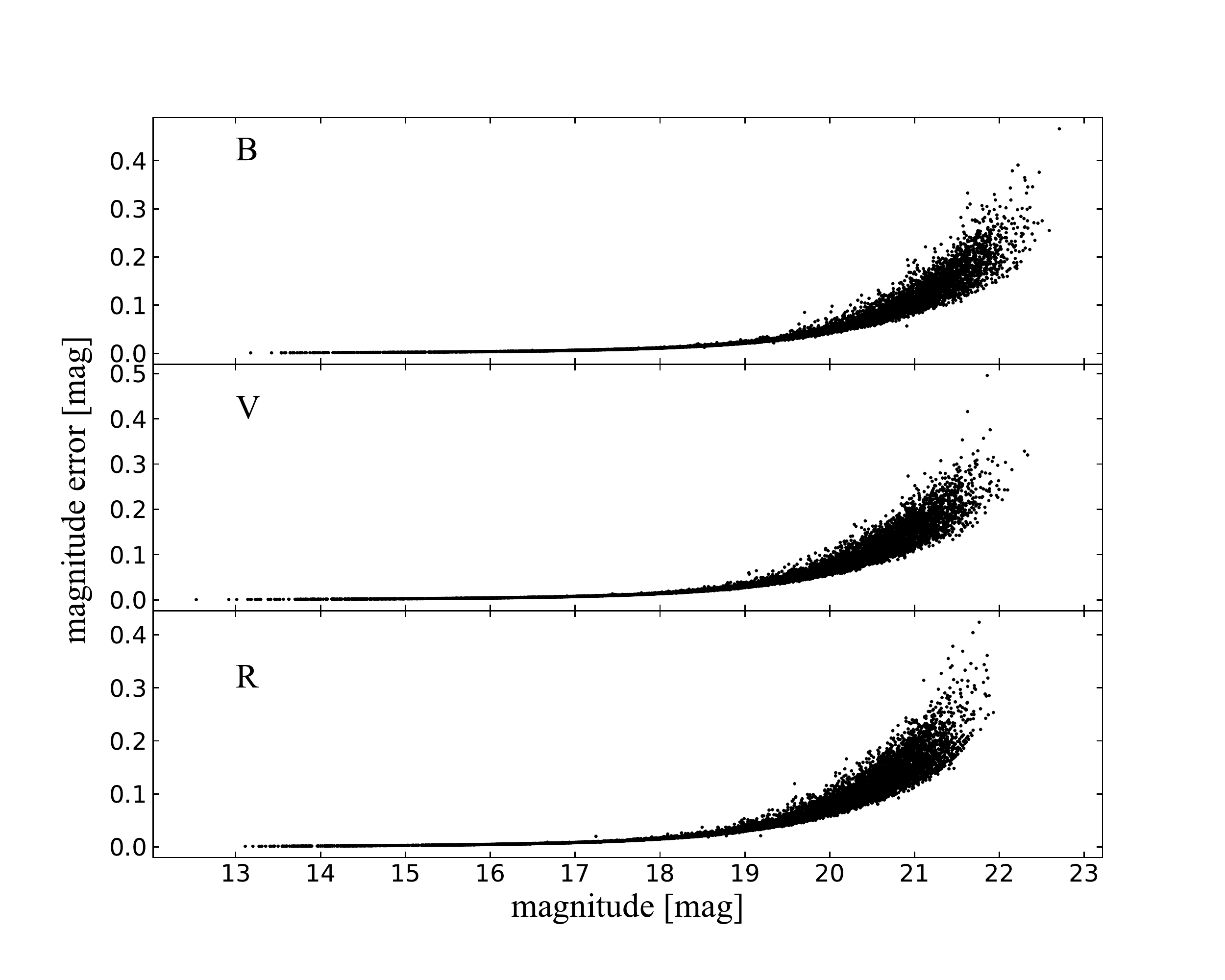}
	\caption{Photometric errors of observations in $B, V$ and $R$ bands.}
	\label{fig:photome-err}
	\end{center}
\end{figure}

Firstly, the CCD frames were reduced with IRAF\footnote{Image Reduction and Analysis Facility is developed and distributed by the National Optical Astronomy Observatory, which is operated by the Association of Universities for Research in Astronomy in Tucson, Arizona, USA, under operative agreement with the National Science Foundation}, including corrections for bias level and flat field. 
We did not consider the dark correction since the dark current is less than $1e^{-}pix^{-1}h^{-1}$ at $-120^{\circ}$C.
To identify objects in the CCD frames, the pixel coordinates of the frames were converted into equatorial coordinates by matching with the third US Naval Observatory CCD Astrograph Catalog (UCAC3).
For the reduced CCD frames, the photometry was carried out by SExtractor\footnote{\url{https://www.astromatic.net/software/sextractor}} \citep{Bertin+Arnouts+1996}.
SExtractor is a program that can perform reasonably well on astronomical images of moderately crowded star fields such as open clusters.
It uses the K-$\sigma$ Clipping method to calculate the distribution of background values with the location of all small areas in CCD frames.
For crowded star fields, it performs median filtering on the distribution in order to suppress possible local overestimations due to bright stars \citep{Bertin+Arnouts+1996}.
Figure~\ref{fig:photome-err} represents the photometric errors of our observations as a function of magnitudes in $B, V$ and $R$ bands.
As illustrated in it, the photometric errors of stars are no more than 0.05 mag in all bands when instrumental magnitudes are smaller than $19.0$ mag.

Then a data processing system of XAO time-domain survey (XAO-pipeline, hereafter) was used to derive the differential light curves of stars that are brighter than $20.0$ mag in instrumental magnitudes \citep{Song+etal+2016, Ma+etal+2018}.
The pipeline is built based on the algorithms provided by \citet{Tamuz+etal+2005, Collier+etal+2006, Ofir+etal+2010} and the IRAF.
The differential magnitude of each star is given by
\begin{equation}
 x_{ij}=m_{ij}-\hat{m}_{j}-\hat{z}_{i}
\label{eq:LebsequeI}
\end{equation}
where $m_{ij}$ is a two-dimensional array of instrumental magnitude, the index $i$ denotes a single CCD frame whine the entire seasons data, the second index $j$ labels an individual star. 
$\hat{m}_{j}$ is the mean instrument magnitude for each numbered star.
$\hat{z}_{i}$ is the zero-point correction for each CCD frame \citep{Collier+etal+2006}.
After repeated iterations, the differential light curves of the stars in the field of view were obtained.

\section{Results} \label{sec: results}
\subsection{Detection of variable stars} \label{sec:detect}
\begin{table*}
\begin{center}
\caption{Parameters of the variable stars in the observed field.\label{tab:allVS}}
\setlength{\tabcolsep}{1pt}
	\begin{tabular}{cccccccccc}
	\hline
	\hline
ID&$\alpha_{2000}$ & $\delta_{2000}$&$V$mag& A & P & Type & PMemb & discovery\\
	&(deg) & (deg) & (mag) & &(d) & & &\\
	\hline
V1&81.942 & +36.042 &16.50 & 0.03(2) & 0.54(5) & $\gamma$Dor &0.8&J09\\
V2&81.902 & +36.129 &16.48 & 0.04(5) & 0.60(8) & $\gamma$Dor &0.5&J09\\
V3&82.276 & +35.577 &16.42 & 0.02(1)& 0.57(4) & $\gamma$Dor &0.5&this work\\
V4&82.057 & +35.865 &16.48 & 0.01(9) & 0.45(2) & $\gamma$Dor &1.0& this work\\
V5&82.252 & +35.794 &16.37 & 0.01(6) & 0.49(8) & $\gamma$Dor &0.8& this work \\
V6&81.776 & +35.964 &16.35 & 0.03(4) & 0.60(5) & $\gamma$Dor &-- &this work\\
V7&81.976 & +36.155 &15.00 & 0.03(3) & 0.31(6) & $\gamma$Dor&-- &this work\\
V8&82.065 & +36.004 &16.56 & 0.06(0) & 0.20(2) & $\delta$Sct&-- &S06\\
V9&82.220 & +35.985 &16.05 & 0.01(2) & 0.11(2) & $\delta$Sct&-- &S06\\
V10&81.826 & +35.632 &14.67 & 0.01(2) & 0.10(7) & $\delta$Sct&-- &J09\\
V11&82.465 & +36.042 &15.31 & 0.00(6) &0.05(0) &$\delta$Sct&-- &J09\\
V12&82.514 & +35.587 &14.80 &0.00(6) & 0.05(8) &$\delta$Sct&-- &J09\\
V13&81.974 & +35.764 &15.88 & 0.00(5) &0.08(3) &$\delta$Sct&-- &J09\\
V14&82.173 & +36.020 &16.19 &0.00(6) & 0.04(6) & $\delta$Sct&0.5 &J09\\
V15&82.236 & +35.939 &17.87 & 0.03(3) & 0.11(2) & $\delta$Sct&-- &this work\\
V16&82.059 & +35.591 &19.00 & 0.35(3)  & 0.38(4) & EW  &-- &S06\\
V17&82.012 & +35.536 &19.57 & 0.40(5)   & 0.34(1)   & EW &-- &S06\\
V18&81.964 & +35.722 &17.11 & 0.32(5)  & 0.46(6) & EW  &-- &S06\\
V19&82.299 & +35.722 &18.16 & 0.40(4)  & 0.38(3)  & EW  &-- &this work\\
V20&81.728 & +36.044 &17.64 & 0.90(7)  & 2.56(6)  & EA  &-- &this work\\
V21&82.250 & +35.589 &18.35 &  1.20(5)  & 1.24(0)  & EA  &-- &this work\\
V22&82.261 & +35.637 &18.62 & 1.50(7)   & 0.58(9)  & EA  &-- &this work\\
V23&81.830 & +35.465 &15.12 & 0.01(1)&0.56(4) &$\gamma$Dor &-- &this work\\
V24&82.072 & +35.820 &17.49 & $>0.1$&-- &unknown &-- &S06\\
		\hline
	\end{tabular}
	\tablecomments{0.86\textwidth}{Column 1: variable stars' ID. Column 2 and 3: right ascension and declination (J2000). Column 4: the mean instrumental magnitude of variable stars in $V$ band, the typical uncertaintices are no more than 0.05 mag. Column 5: the amplitude of folded light curves in $V$ band. Column 6: the main period of variable stars calculated by LombScargle. The last-digit errors of the amplitudes and periods are given in parentheses. Column 7: type of variable stars. Column 8: membership probability of variable stars provided by \citet{Cantat-Gaudin+etal+2018}. Column 9: reference of variables detection. S06 and J09 indicate that the variable stars were found by \cite{Szabo+etal+2006} and \cite{Jeon+2009}, respectively.}
\end{center}	
\end{table*}

\begin{figure}[htp]
\begin{center}	
\includegraphics[height=12cm,width=9cm,clip,angle=0]{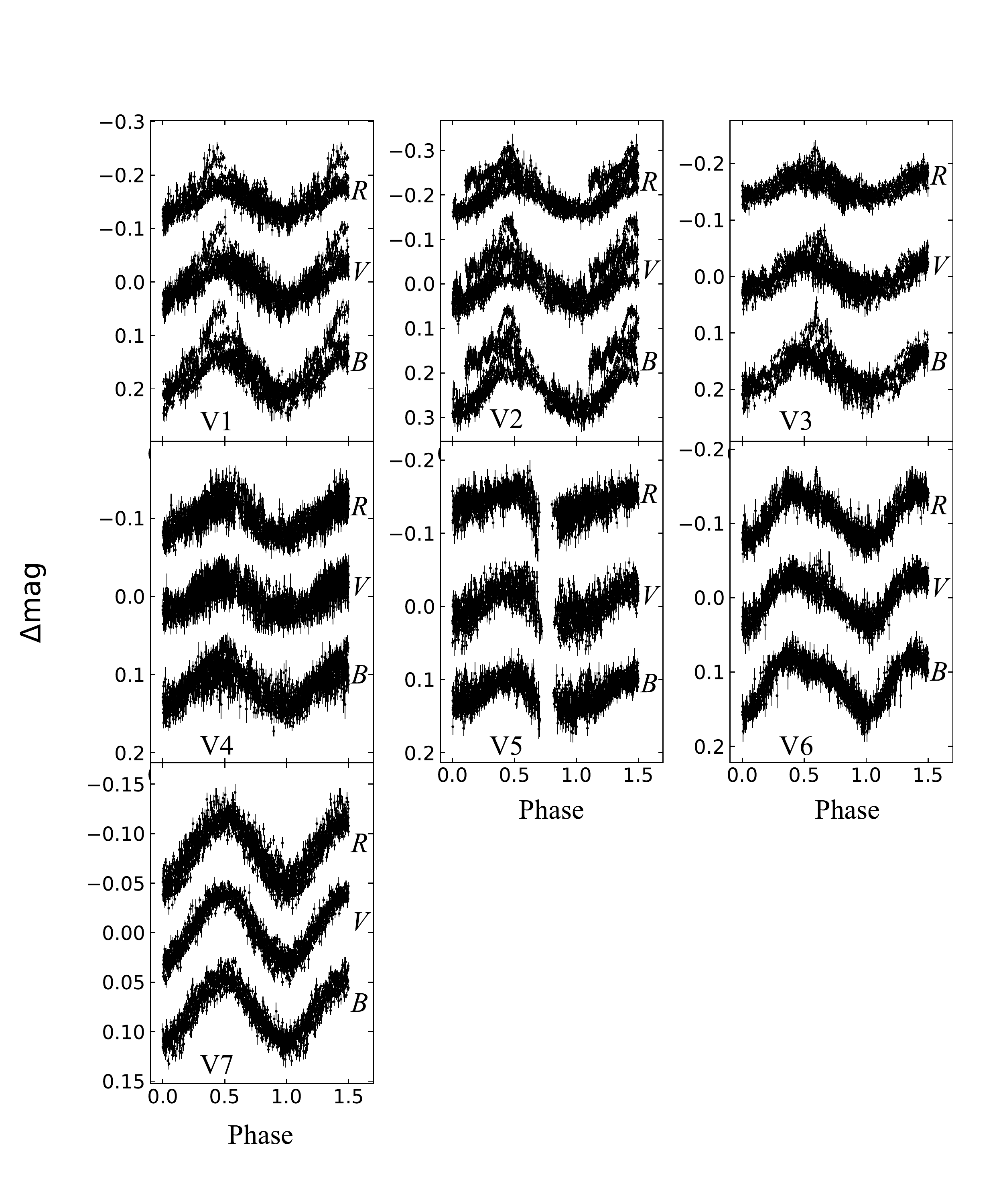}
\caption{$R, V$ and $B$ bands phase-folded light curves for star V1 to V7 listed in Table~\ref{tab:allVS}.}
\label{fig:folded-dor}	
\end{center}
\end{figure}

\begin{figure}[htp]
\begin{center}
\includegraphics[height=12cm,width=9cm,clip,angle=0]{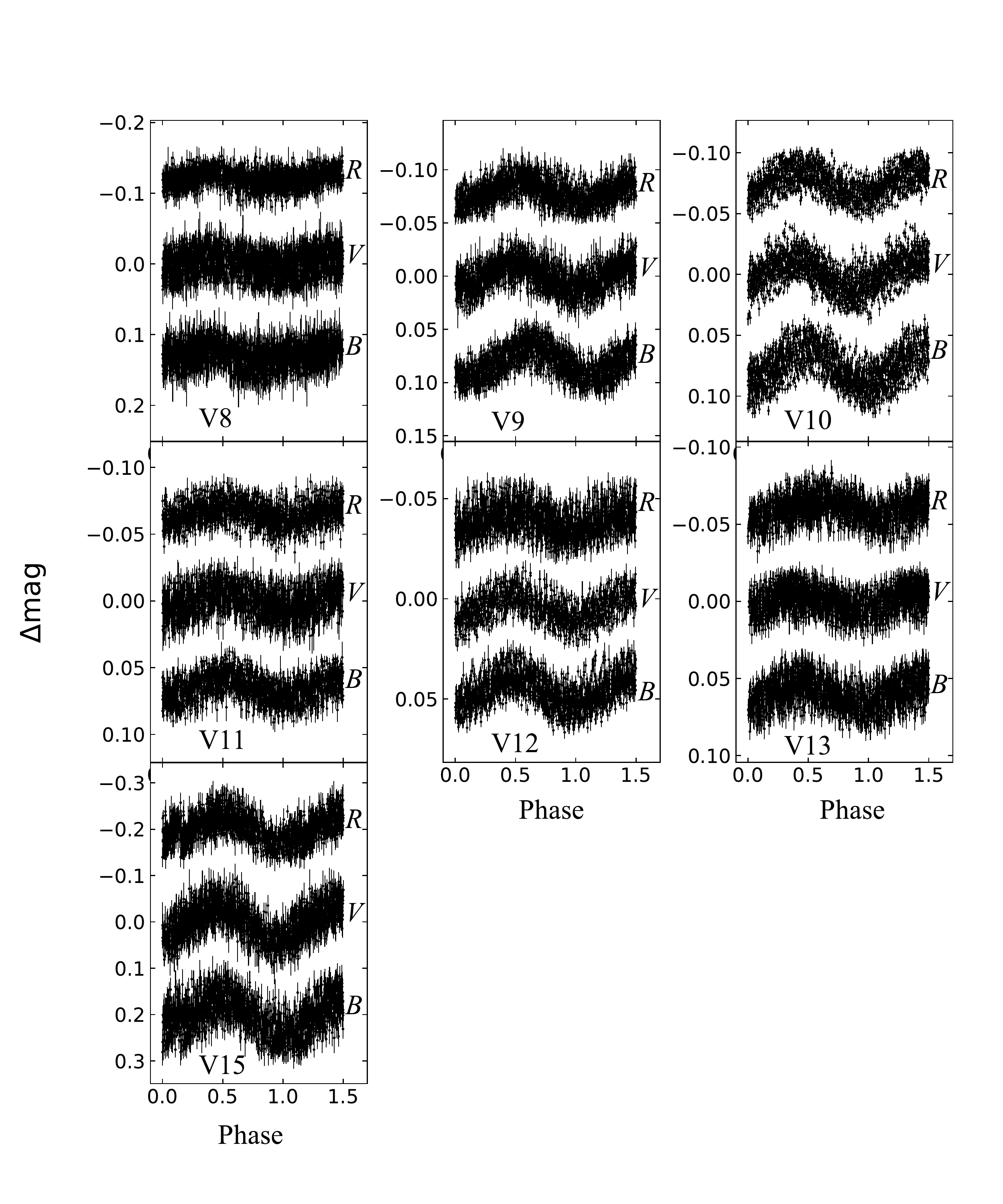}
\caption{Same as Fig.~\ref{fig:folded-dor}, but for star V8 to V13 and V15 listed in Table~\ref{tab:allVS}.}
\label{fig:folded-Sct}
\end{center}
\end{figure}

\begin{figure}[htp]
\begin{center}	
\includegraphics[height=12cm,width=9cm,clip,angle=0]{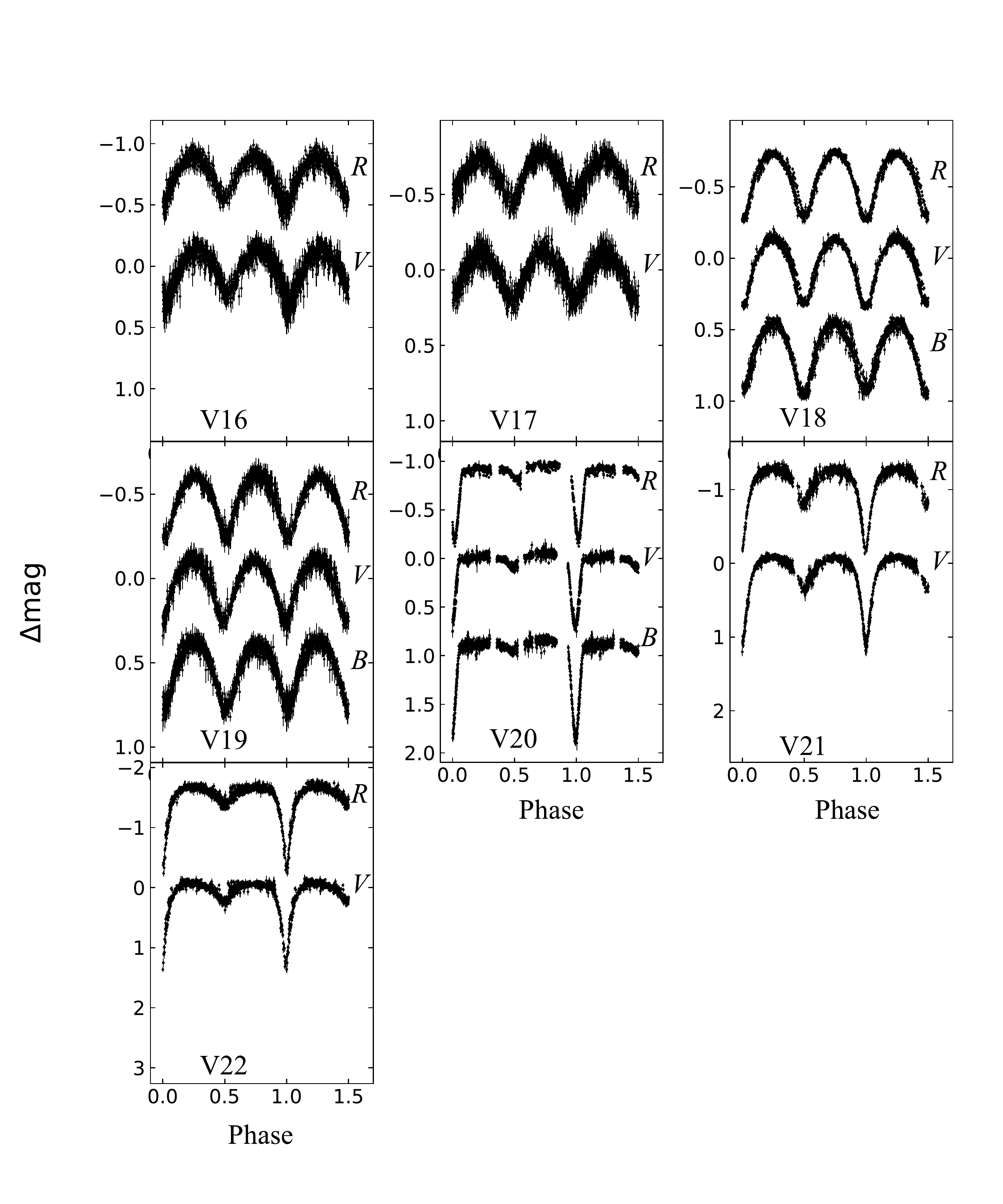}		
	\caption{Same as Fig.~\ref{fig:folded-dor}, but for star V16 to V22 listed in Table~\ref{tab:allVS}. The $B$ band phase-folded light curves of V16, V20, V21 and V22 are absent because the mean instrumental magnitudes of these stars in $B$ band are larger than 19 mag.}
\label{fig:folded-eb}
\end{center}
\end{figure}

\begin{figure}[htp]
\begin{center}
\includegraphics[height=6cm,width=12cm,clip,angle=0]{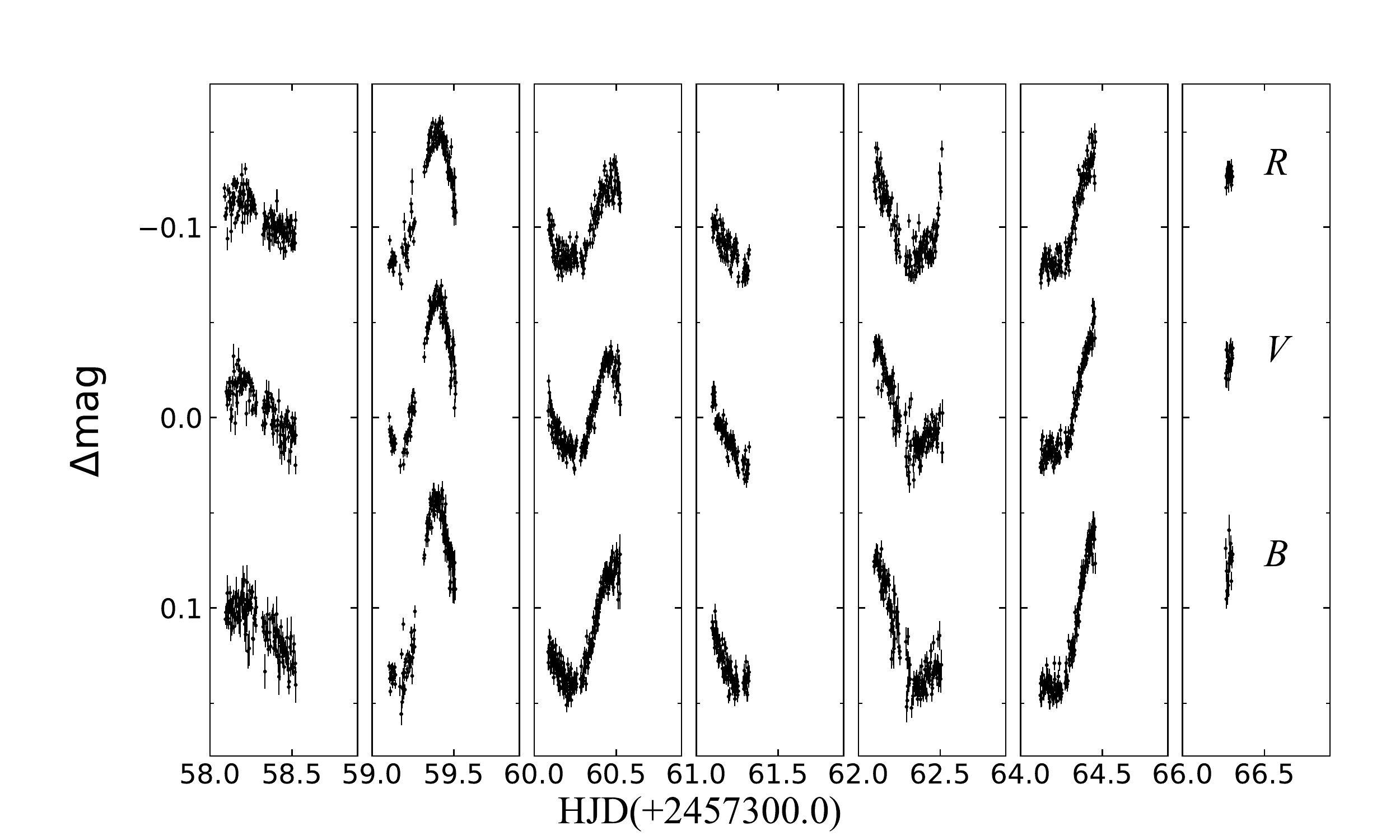}	
	\caption{Light curves of V23 in $R, V$ and $B$ bands.}
\label{fig:V22HJD}
\end{center}
\end{figure}

\begin{figure}[h]
   \begin{center}
	\includegraphics[height=6cm,width=6cm,clip,angle=0]{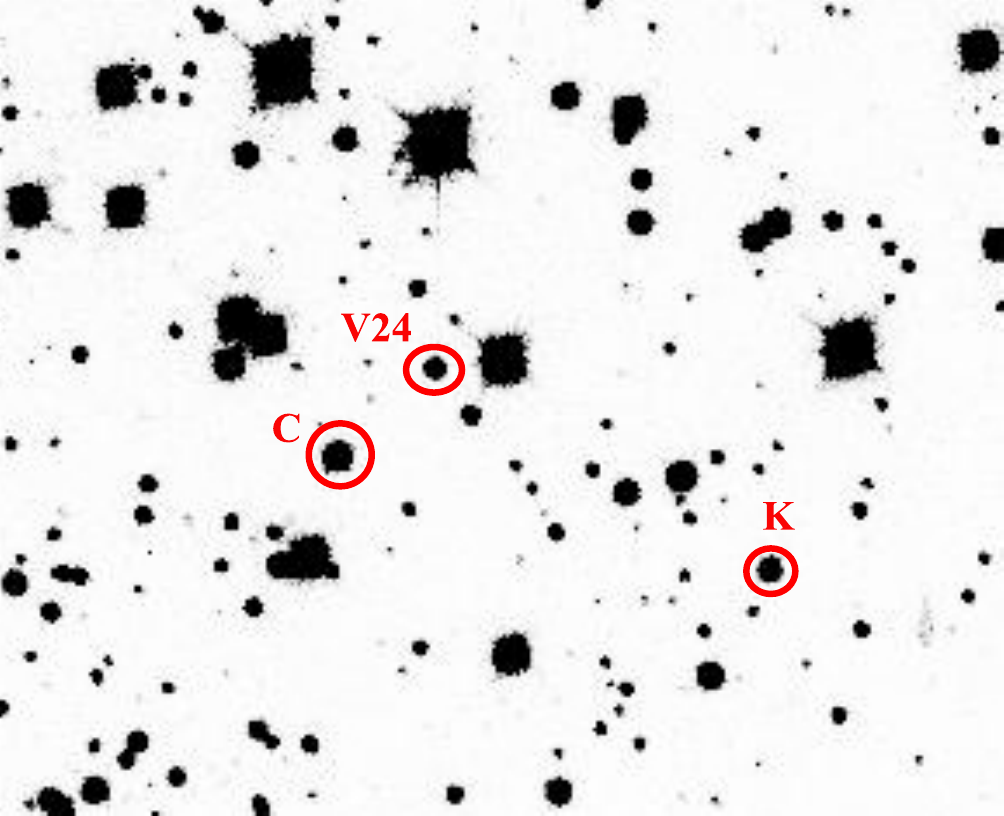}
	\caption{V24, the comparison star and the check star are marked with red circles. They are marked with characters 'V24', 'C' and 'K' respectively.}
	\label{fig:TCK}
	\end{center}
\end{figure}

\begin{figure}[htp]
\begin{center}
\includegraphics[height=15cm,width=8cm,clip,angle=0]{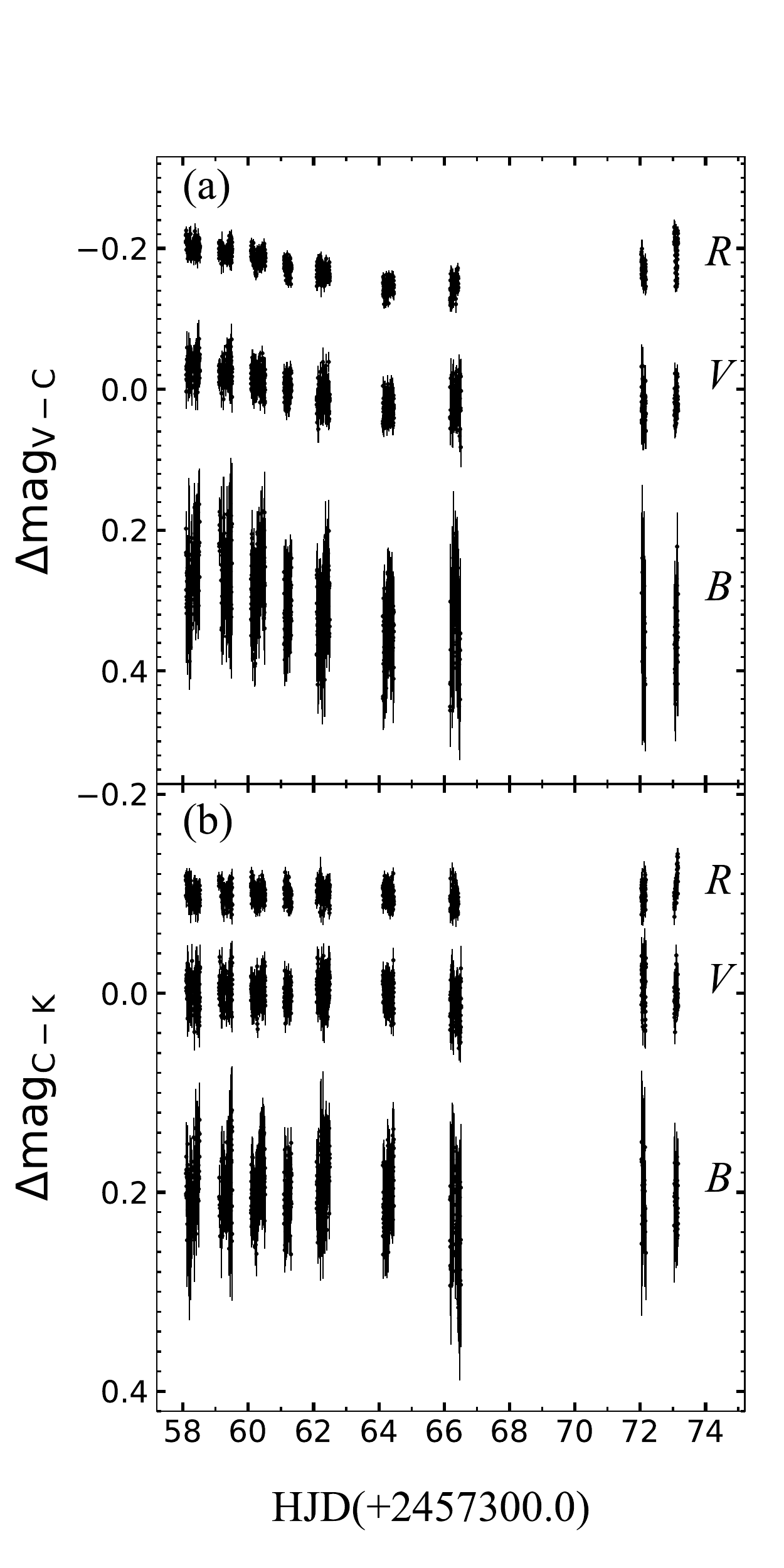}
	\caption{Light curves of V24 in $R, V$ and $B$ bands. (a): Magnitude differences of V24 (V) and comparison star (C). (b): Magnitude differences of comparison (C) and check star (K).}
\label{fig:LPHJD}
\end{center}
\end{figure}

As listed in Table~\ref{tab:allVS}, a total of 24 variable stars were detected in this work.
The following two methods were used to search for variable stars.
First, we visually inspect all the light curves to search for eclipse binaries or long-term variable stars.
Stars can be considered as variable stars when their light curves show significant simultaneous variations in $B, V$, and $R$ bands.
Therefore, 23 variable stars with significant light curve variations were selected.
We calculated the most significant periods of these stars using LombScargle periodogram  \citep{VanderPlas+etal+2012,VanderPlas+etal+2015}, which was achieved by the sub-package of Astropy \citep{Astropy+etal+2013}.
The folded light curves of 21 periodic stars are presented in Fig.s~\ref{fig:folded-dor}, \ref{fig:folded-Sct}, and \ref{fig:folded-eb}.
Figure~\ref{fig:V22HJD} shows the light curves of V23, where we see a significant variation.
However, we can not find a reliable period to fold the light curves.
To verify the long-term and small-amplitude variation of V24 detected through XAO-pipeline, we chose a comparison star and a check star to perform differential photometry on V24.
The parameters of the comparison star and the check star are listed in Table~\ref{tab:ROC}.
Figure~\ref{fig:TCK} shows their positions on the CCD frame.
Figure~\ref{fig:LPHJD} shows their differential light curves.
The light curves of V24 and the comparison star are consistent with that obtained through XAO-pipeline.

\begin{table*}
\begin{center}	
	\caption{Coordinates and visual magnitude of the target, comparison and check stars.\label{tab:ROC}}
	\setlength{\tabcolsep}{2.5pt}
	\begin{tabular}{ccccccc}
		\hline
		\hline
		Stars & $\alpha_{2000}$ & $\delta_{2000}$ & $V$ & $B$ & $B-V$ \\
		 &(deg) & (deg) & (mag) & (mag) &($\Delta$ mag )\\
		\hline
V24 & 82.072 &  +35.820 & $17.498\pm0.011$ & $18.959\pm0.032$  &$1.461\pm0.023$ \\
Comparison star (C)  &  82.081 & +35.829 &$15.97\pm0.004$ & $17.138\pm0.007$ &$1.168\pm0.005$ \\
Check star (K) &  82.031 & +35.832 &$16.674\pm0.006$ &$17.814\pm0.010$ &$1.140\pm0.008$  \\
		\hline
	\end{tabular}
\end{center}
\end{table*}

\begin{figure}[h]
   \centering
   \includegraphics[height=12cm,width=9cm,clip,angle=0]{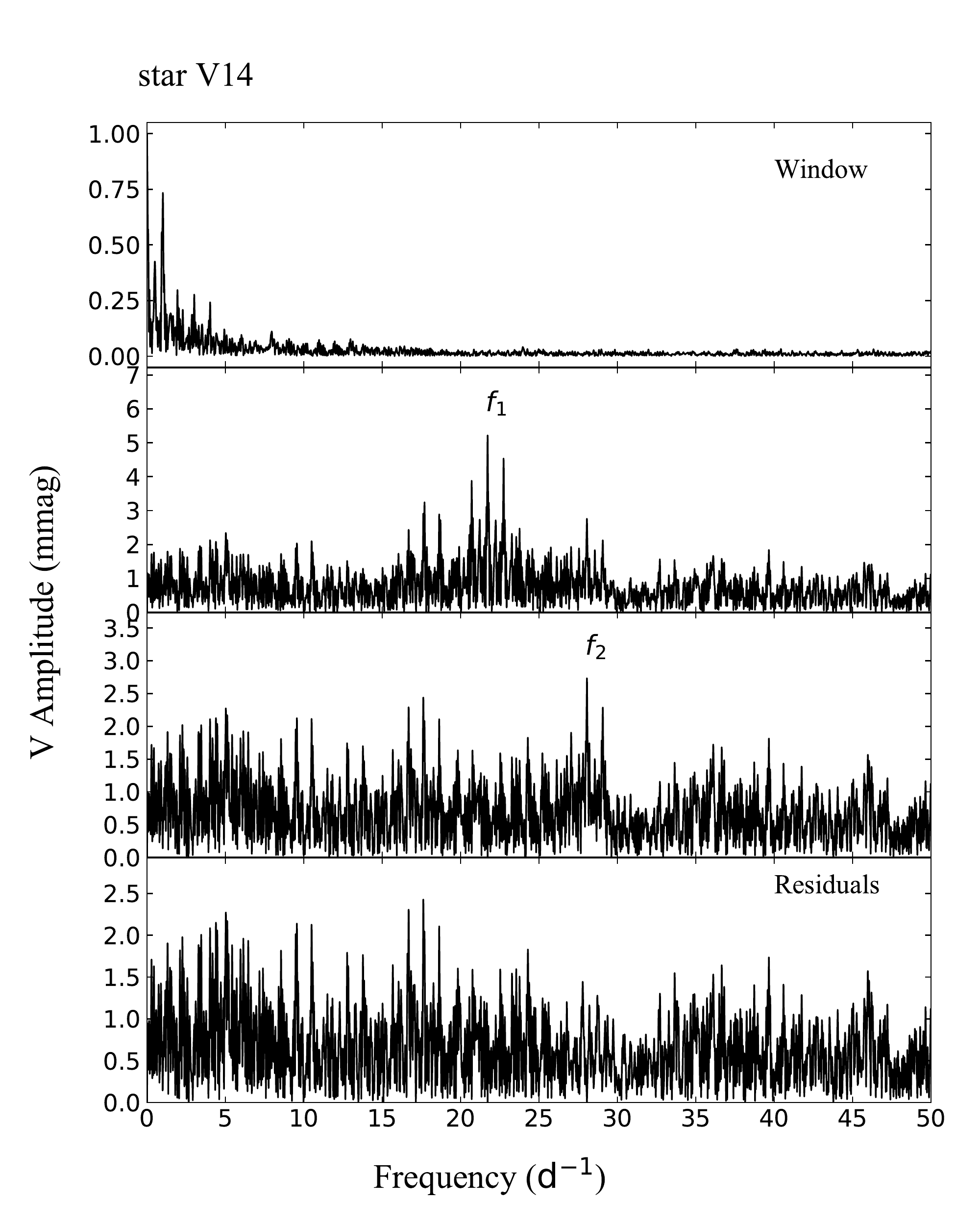}
   \caption{Frequency analysis for star V14. The top panel: the spectral window of the light curve of V24 in $V$ band. The sub-sequent panels: amplitude spectra of V14 with successive pre-whitening.}
   \label{fig:as-0915}
   \end{figure}

To detect the missing periodic variable stars, we then performed frequency analyses for all other stars' time-series data in $V$ band by using Period04 \citep{Lenz+Breger+2004, Lenz+Breger+2005}.
This software adopts single-frequency Fourier and multifrequency nonlinear least-squares fitting algorithms.
It offers tools to extract the individual frequencies from the multi-periodic content of time series and provides a flexible interface to perform multiple-frequency fits \citep{Lenz+Breger+2004}.
The frequencies of the intrinsic and statistically significant peaks in the Fourier spectra can be extracted via iterative pre-whitening.
The frequency investigations were stopped when the signal-to-noise (S/N) value is less than 4.0 \citep{Breger+etal+1993}.
Amplitude spectra were computed after pre-whitening.
The amplitude spectra of these stars are regarded as another criterion for finding variable stars.
In this way, we detected one more variable star, whose star ID is V14.
The amplitude spectrum of this variable star is shown in Fig.~\ref{fig:as-0915}.
Table~\ref{tab:fa-V1-V24} lists the frequency solutions, amplitudes, and S/N values of the variable star.

\subsection{Cluster membership of the variable stars }\label{sec:membership}
For variable stars detected in this work, we used the cluster members established by \citet{Cantat-Gaudin+etal+2018} to identify their membership.
\citet{Cantat-Gaudin+etal+2018} used the membership assignment code Unsupervised Photometric Membership Assignment in Stellar Clusters (UPMASK) and {\it Gaia} DR2, and they identified 807 members of NGC 1912.
Considering the precision of the data, the sources they used are $G < 18$ mag.
The membership probabilities of these stars were obtained after 10 iterations of UPMASK, from 0 to 100\% by an increment of 10\%.
Among 807 cluster members, there are 474 stars with membership probabilities over 50\%.
To get the cluster memberships for detected variable stars, we matched the equatorial coordinates of 2908 stars we observed with that of the 807 reported members within $3^{''}$.
As a result, 514 cluster members are detected in our field of view.
Six of the 514 cluster members are variable stars studied in this work.
The membership probabilities of these variable stars V1 to V5 and V14 are presented in column 8 of Table~\ref{tab:allVS}.
All these membership probabilities are larger than 50\% (see details in Table~\ref{tab:allVS}).

\subsection{Color-magnitude diagram} \label{sec:CMD}
\begin{figure}
\begin{center} 
\includegraphics[width=1\linewidth]{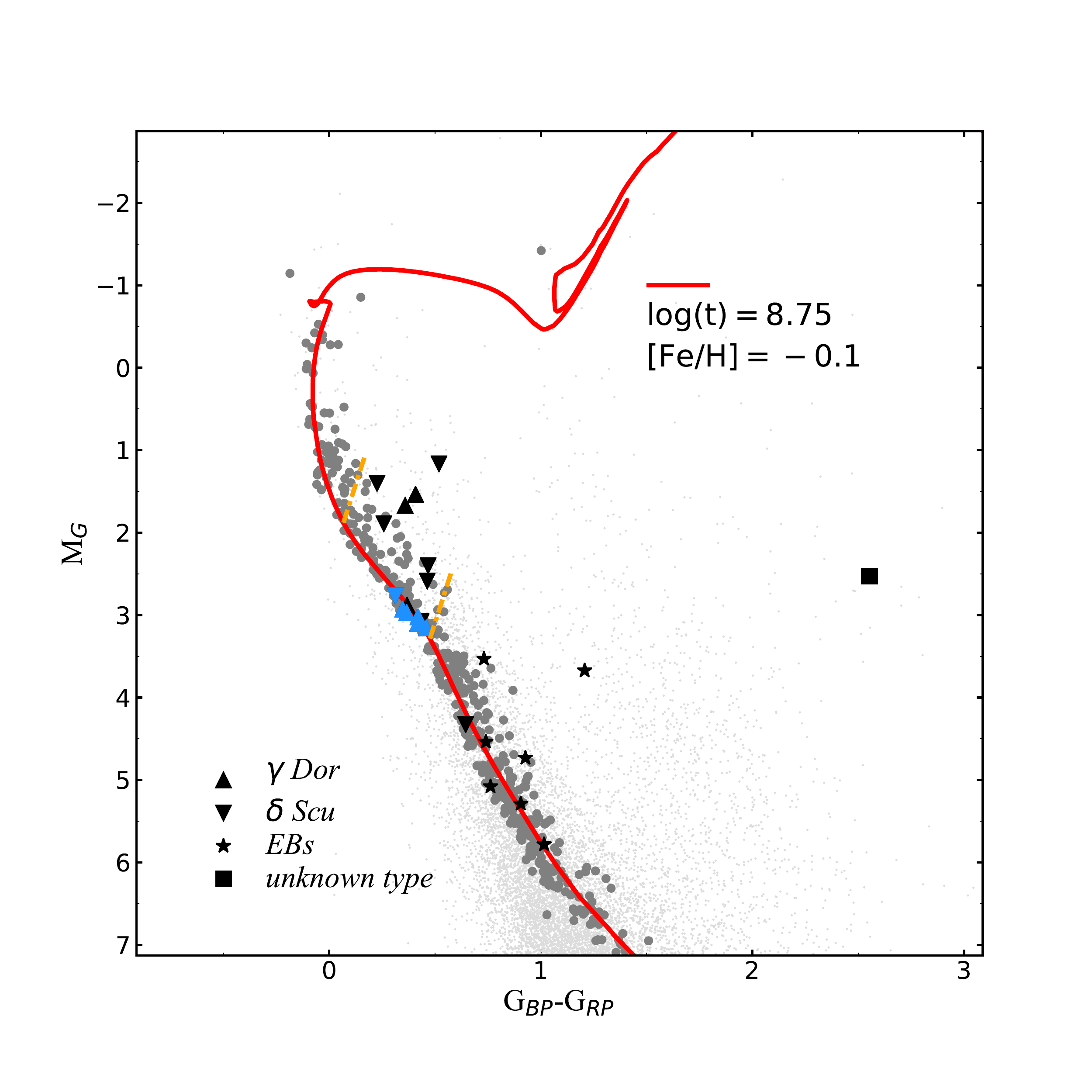}            
	\caption{CMD of NGC 1912 with isochrone and the instability strip of $\gamma$\,Dor stars. Grey dots represent stars with membership probabilities over 50\%. The isochrone is indicated by the red line(see text for details). The approximate range of theoretical $\gamma$\,Dor star instability strip on the isochronous line is pointed by two orange dashed lines. Different types of variable stars are plotted with different symbols in the diagram. The corresponding legend is at the bottom left of the diagram. Blue symbols: they are variable stars detected in this work and were identified as cluster members by \citet{Cantat-Gaudin+etal+2018}. Black symbols: they are variable stars but were not reported as cluster members by \citet{Cantat-Gaudin+etal+2018}.}
\label{fig:member-CMD} 
\end{center}
\end{figure}

The main physical parameters of the six variable cluster members can be estimated by an isochrone fitting of color-magnitude diagram (CMD).
Figure~\ref{fig:member-CMD} shows the isochrone fitting results for the cluster members with membership probabilities over 50\%.
The theoretical isochrone was derived from stellar evolutionary tracks computed with PARSEC code \citep{Bressan+etal+2012} and its initial parameters were adopted from \cite{Monteiro+Dias+2019} and \citet{Reddy+etal+2015} for surface gravity $log(age)=8.75$ and metallicity $[Fe/H]=-0.1$.
For each star plotted in the CMD, their observational magnitudes $G$ mag and colors $G_{BP}$ - $G_{RP}$ were transformed into absolute magnitude M$_{G}$ and intrinsic colors $(G_{BP}-G_{RP})_{0}$.
We used distance modulus $m-M=10.03$ mag, extinction A$_{G}$ = 0.841, and colour excess E($BP-RP$) = 0.41.
The distance modulus was derived from the distance of cluster $d=1014$ pc \citep{Monteiro+Dias+2019}.
Referring to \citet{Monteiro+Dias+2019}, we know that the colour excess of NGC 1912 in $B-V$ color is E($B-V$) = 0.307.
The attenuation for magnitude M$_{G}$ can be calculated as:
\begin{equation}
A_{G} = R_{G} \times E(B-V)
\label{eq:extinction}
\end{equation}

The excess in $G_{BP}$ - $G_{RP}$ color is calculated as:
\begin{equation}
E(BP-RP) = (R_{BP}-R_{RP}) \times E(B-V)
\label{eq:redden}
\end{equation}

Where the extinction coefficients for $Gaia$ filters are R$_{BP}$ = 3.374, R$_{RP}$ = 2.035, and R$_{G}$ = 2.740 \citep{Casagrande+VandenBerg+2018}. 

Referring to \citet{Dupret+etal+2005a, Dupret+etal+2005b}, \citet{Aerts+etal+2010} provided a approximate range of effective temperature $log T_{eff}$ and $log(L/L_{\odot})$ for the instability strip of $\gamma$\,Dor stars, which are $log T_{eff} \in [3.83,3.90]$ and $log(L/L_{\odot}) \in [0.7,1.1]$.
We showed the the instability strip of $\gamma$\,Dor stars in the isochrone, and plotted the 24 detected variable stars in Fig.~\ref{fig:member-CMD}.
These variable stars are shown as different symbols based on their types, which are identified in Section \ref{sec:membership}.
As shown in Fig.~\ref{fig:member-CMD}, the six variable members are located in the main sequence, and they are all within the instability strip of $\gamma$\,Dor stars that we marked.

\subsection{Classification of variable stars}\label{sec:class}
 
We classified the variable stars mainly by their main periods and the locations of variable cluster members in the CMD.
We performed frequency analysis for the variable stars (except eclipse binaries).
The corresponding frequency solutions, amplitudes, and S/N values in the $V$ band of these stars are listed in Table~\ref{tab:fa-V1-V24}.
The amplitude spectra of these variable stars are shown in Fig.~\ref{fig:as-0915}, Fig.~\ref{fig:as-V1-V4} to Fig.~\ref{fig:as-V23}.

We followed the criteria of \citet{Grigahc_ne_2010} and \citet{Uytterhoeven2011T} to classify these variable stars as $\gamma$\,Dor or $\delta$\,Sct stars.
The distinction between $\gamma$\,Dor and $\delta$\,Sct stars is based completely on whether their frequencies of peaks with significant amplitudes are less or large than $5\,\mathrm{d^{-1}}$.
From Table~\ref{tab:fa-V1-V24}, the dominant frequencies of V1 to V7 and V23 are less than $5\,\mathrm{d^{-1}}$.
At the same time, the variable cluster members (V1 to V5) are located whine the instability strip of $\gamma$\,Dor stars in the CMD of the open cluster NGC 1912.
While the dominant frequencies of V8 to V15 are around or large than $5\,\mathrm{d^{-1}}$.
We can distinguish that V1 to V7 and V23 are $\gamma$\,Dor stars, and V8 to V15 are $\delta$\,Sct stars. 

V16...V22: Figure~\ref{fig:folded-eb} shows their folded light curves.
V16, V17, V21 and V22 are too faint in $B$ band to perform photometric analysis so that their folded light curves are empty in this band.
From the shapes of these light curves and their periods, we can identify V15...V18 and V19...V21 are EW type and EA type stars respectively.

V24: From Fig.~\ref{fig:LPHJD}, we can observe the long-term variabilities of light curves in three bands.
The amplitude spectrum of the star is shown in Fig.~\ref{fig:as-V24}.
As listed in Table~\ref{tab:fa-V1-V24}, we extract two frequencies from the photometric data of V24.
The main period of the star is about four days.
However, the light curves of V24 show a long-term variation longer than 4 days.
The position of V24 on the CMD deviates from the main sequence.
Thus, the determination of the type of this variable star and the two frequencies we extracted may require more and longer observations to confirm.

\section{Discussion} \label{sec:discu}

\subsection{Comparison with previous work}

Variable stars in NGC 1912 were reported by \cite{Szabo+etal+2006} and \cite{Jeon+2009}.
\cite{Szabo+etal+2006} detected 14 variable stars in the field of NGC 1912 using a 60/90/180 cm Schmidt telescope with angular resolution of $1.1^{''}$/pixel. 
The 14 variables were classified as one EW type stars, three EA type stars, five $\delta$\,Sct pulsators, four long-term variable stars, and one unknown type variable star by them.
In a $1.0^{\circ} \times 1.5^{\circ}$ field of view around the center of NGC 1912, \cite{Jeon+2009} newly detected 15 $\delta$\,Sct stars and two $\gamma$\,Dor stars, and they confirmed three $\delta$\,Sct stars which were reported by \cite{Szabo+etal+2006}.
And 14 of the 20 detected stars were located within radius $30^{'}$ from the center of NGC 1912.
The time-series CCD images for 23 nights they obtained were taken through a small refracting telescope (D = 155 mm, f = 1050 mm).

In our field of view, we observed 24 variable stars, among which 13 variable stars are previous known.
Two $\delta$\,Sct stars (V8, V9), three EW type stars ( V16...V18) and V24 were reported by \cite{Szabo+etal+2006}.
Two $\gamma$\,Dor stars (V1 and V2) and five $\delta$\,Sct stars (V10... V14) were found by \cite{Jeon+2009}, and another $\delta$\,Sct stars (V9) was confirmed by them.
The parameters of these known variables listed in Table~\ref{tab:allVS} are similar to that provided by the previous works.

 \cite{Jeon+2009} performed frequency analysis on their detected pulsating stars by using Period04.
The dominant frequencies of these pulsators they provided are consistent with that we list in Table~\ref{tab:fa-V1-V24}.
However, there are some differences in other oscillation frequencies of V1, V2, V10 and V14.
The second and third oscillation frequencies of V1 and V10 are different between the previous and the present work.
The second and third oscillation frequencies of V1 are $1.6160\pm0.0022\,\mathrm{d^{-1}}$ and $1.5100\pm0.0026\,\mathrm{d^{-1}}$ respectively in this work, while they were $1.937\,\mathrm{d^{-1}}$ and $1.654\,\mathrm{d^{-1}}$ respectively in the previous work.
The second and third oscillation frequencies of V10 are $8.4972\pm0.0029\,\mathrm{d^{-1}}$ and $10.6937\pm0.0027\,\mathrm{d^{-1}}$ in the present work, however, they are $9.7021\,\mathrm{d^{-1}}$ and $7.596\,\mathrm{d^{-1}}$ in \cite{Jeon+2009} respectively.
Moreover, the frequencies we extracted from the time-series data of V10 and V14 are one more frequency than the previous work.
Two frequencies of V2 were previously found, but the second oscillation frequency $2.310\,\mathrm{d^{-1}}$ can not be identified in this work.

\begin{table}
	\begin{center}
		\caption{Frequency solutions for three previous known variable stars.\label{tab:fa-C1-C3}}
		\setlength{\tabcolsep}{2.5pt}
		\begin{tabular}{ccccccc}
		\hline
		\hline
		star ID &$\alpha_{2000}$ & $\delta_{2000}$&No. & $F_{19}$ & $A_{19}$ & $S/N_{19}$ \\
			&(deg) & (deg) & &(c/d) &(mag) & \\
		 \hline
C1&81.9772&35.8637&$f_{1}$&$0.9522\pm0.0027$ &$0.1334\pm0.0099$&6.6\\
C2&82.2203&35.5810&$f_{1}$&$0.9091\pm0.0015$ &$0.0063\pm0.0004$&14.0\\
 & & &$f_{2}$&$0.5109\pm0.0021$ &$0.0057\pm0.0004$&10.0\\
C3&82.2532&35.8946&--&-- &--&$< 4.0$\\
\hline	 
\end{tabular}
	\tablecomments{0.86\textwidth}{The same as Table~\ref{tab:fa-V1-V24} but for three previous known variable stars C1 to C3.}
	\end{center}
\end{table}

\begin{figure}[h]
   \centering
   \includegraphics[height=10cm,width=9cm,clip,angle=00]{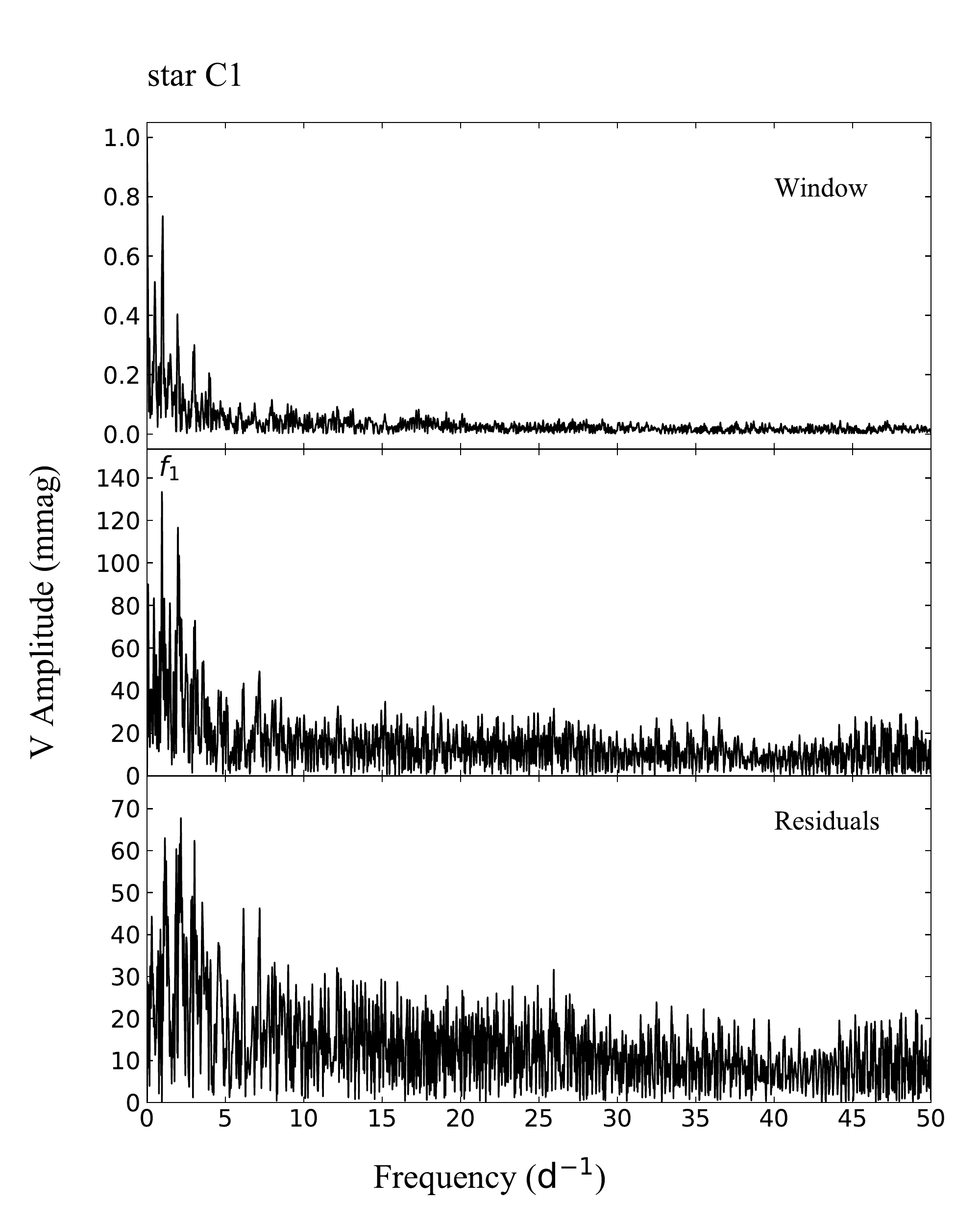}
   \caption{The same as Fig.~\ref{fig:as-0915} but for star C1.}
   \label{fig:as-C1}
   \end{figure}
   
    \begin{figure}[h]
   \centering
   \includegraphics[height=14cm,width=9cm,clip,angle=0]{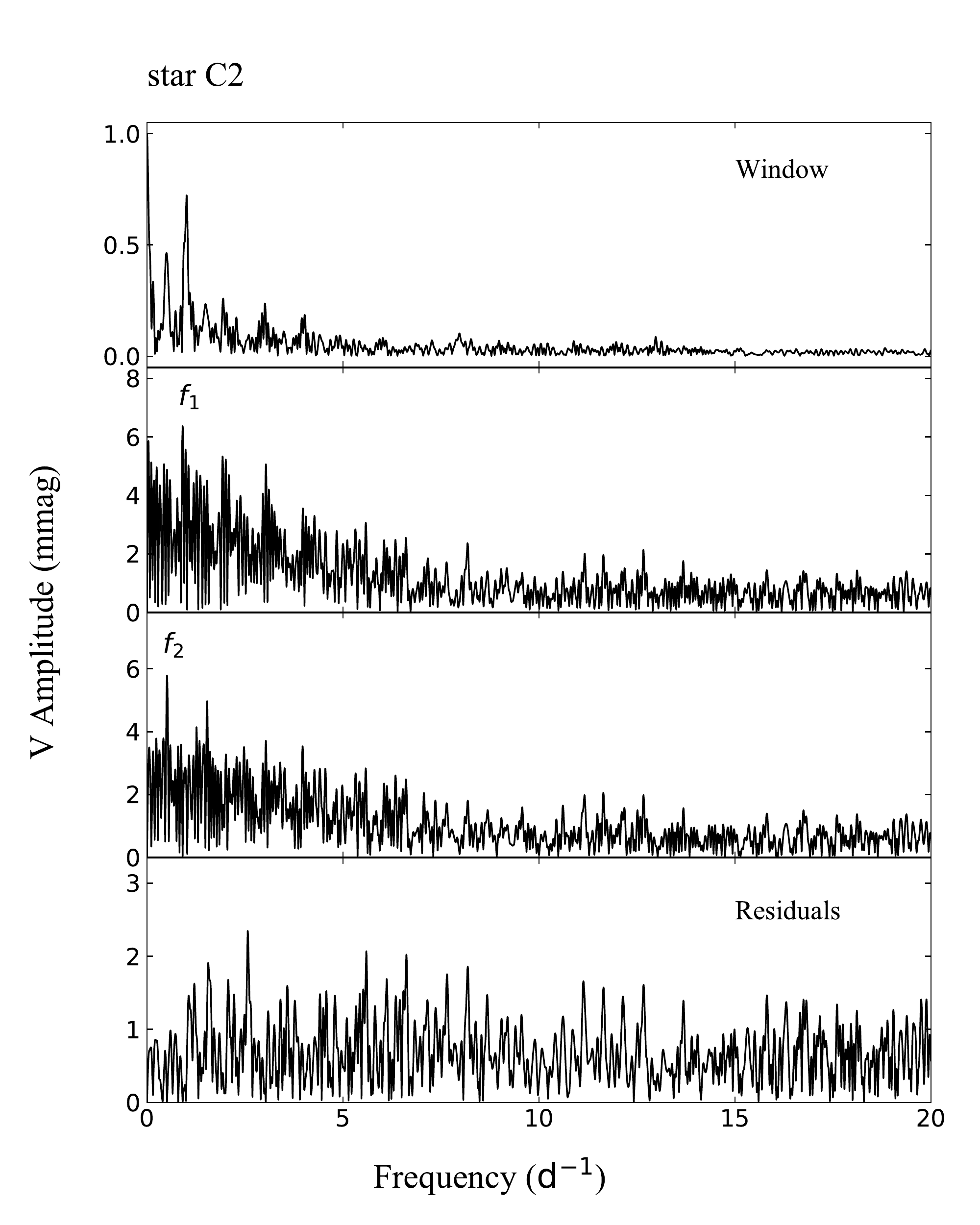}
   \caption{The same as Fig.~\ref{fig:as-0915} but for star C2.}
   \label{fig:as-C2}
   \end{figure}

Apart from the 24 variable stars, we also observed 11 stars which were in \cite{Szabo+etal+2006} and \cite{Jeon+2009} but can't be confirmed in this work.
Several reasons may lead to the inconsistency.
Three of the 11 stars are too faint or too bright to secure reliable variable detections in this work. 
Five of the 11 stars are close to nearby bright stars, it is hard for NOWT to distinguish them.
For the other three stars (named as C1, C2, and C3), we used the software Period 04 to perform frequency analysis on their time-series data obtained in this work.
The results of the frequency analyses are listed in Table~\ref{tab:fa-C1-C3}.
And their amplitude spectra are shown in Fig.~\ref{fig:as-C1} and Fig.~\ref{fig:as-C2}.
C1 and C2 were considered to be long-term variable stars by \cite{Szabo+etal+2006}, but the possible periods of these stars were not given.
In this study, the extracted frequency of C1 is $0.9522\pm0.0027\,\mathrm{d^{-1}}$.
We detected two independent frequencies from the observation data of C2, which are $0.9091\pm0.0015\,\mathrm{d^{-1}}$ and $0.5109\pm0.0021\,\mathrm{d^{-1}}$.
The corresponding periods are one day and two days within $1\sigma$ error,  respectively.
However, observations in this work are affected by the interruption of daytime, we cannot determine whether these periodic signals are caused by the stars' intrinsic variation or the interval of our observations.
Since the peaks in the amplitude spectrum of C3 are lower than four times the noise level, we do not show its amplitude spectrum.
Additional precise observations may be required to further study these variable stars.

Otherwise,  variables observed by the previous work are outside our field of view.
They are not taken into account in this section.

\subsection{parameters of variable cluster members}\label{sec:para}

\begin{figure}[h]
\begin{center}
	\includegraphics[width=1\linewidth,angle=0]{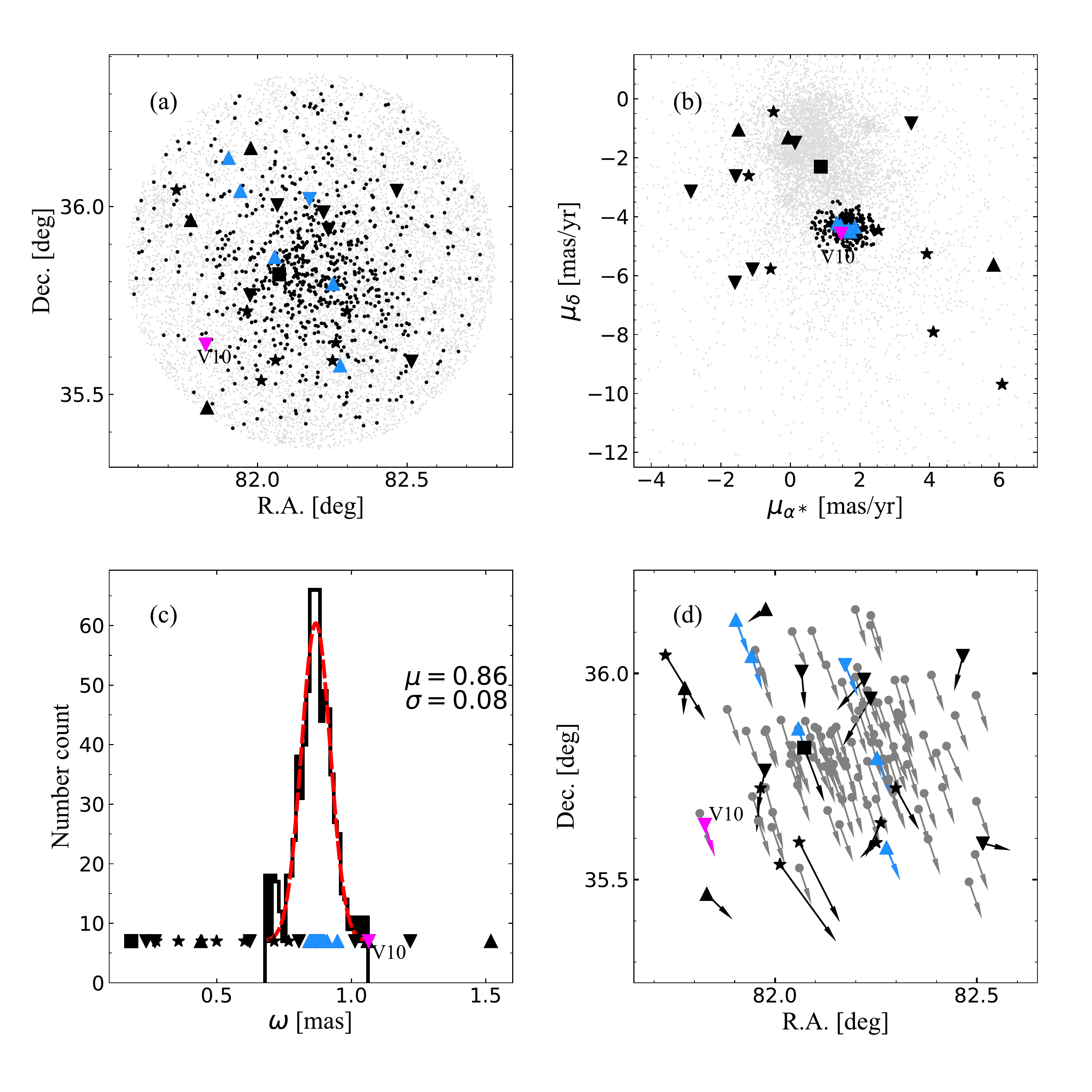}
	\caption{ (a): Spatial distribution for the cluster members and detected variable stars, ($\alpha_{J2000}$ $\delta_{J2000}$). (b): proper motion distribution for the cluster members and detected variable stars.  (c): Histogram of parallax ($\omega$) for cluster members (black line) and variable stars obtained in this work. The histogram of members' parallax can be fitted with a single Gaussian profile(red dashed line). (d): Spatial distribution for the cluster members and detected variable stars with their tangential velocities. Arrows point out the direction of tangential velocities for each star. Arrow length is in proportion to the tangential speed. In top panels, light grey dots mean the complete samples of NGC 1912. Black dots represent all cluster members established by \citet{Cantat-Gaudin+etal+2018}. The meanings of rest marks are identical to that in Fig.~ \ref{fig:member-CMD} but magenta symbol means the probable variable cluster member in Section~\ref{sec:other}. The grey dots in panel (d) represent stars with membership probabilities over 90\%.}
	\label{fig:member-5D}
\end{center}
\end{figure}

Variable stars V1 to V5 and V14 were reported by \citet{Cantat-Gaudin+etal+2018} as cluster members of NGC 1912, since they concentrated within the field of NGC 1912 on the sky and are homogeneous with the other members in proper motions and parallaxes.
Figure~\ref{fig:member-5D} demonstrates the homogeneousness of these variable cluster members in celestial position ($\alpha$, $\delta$), proper motion ($\mu_{\alpha*}$, $\mu_{\delta}$), and parallax ($\omega$).
As shown in the panel (a) of Fig.~\ref{fig:member-5D}, these variable stars are located in the field with center ($\alpha=5^{h}28^{m}42.7^{s}$, $\delta=35^{\circ}51^{'}32^{''}$) and redius $r = 30^{'}$.
The field covers the field of NGC 1912 reported by \citet{Pandey+etal+2007}, center in ($\alpha=5^{h}28^{m}43^{s}$, $\delta=35^{\circ}51^{'}18^{''}$) with $r\backsim14^{'}$.
Cluster members perform uniform bulk motions, they will be over-density in the proper motion diagram.
In the panel (b) of Fig.~\ref{fig:member-5D}, we see that the members of NGC 1912 cluster together showing different proper motions with the other stars.
And the six variable stars perform uniform bulk motions with the members, since their proper motion values are similar to the proper motion of NGC 1912 reported by \cite{Dias+etal+2014} ($\mu_{\alpha*}=-0.52 \pm 1.44$ mas/yr and $\mu_{\delta}=-4.14\pm 1.54$ mas/yr).
As we can see in the panel (c) of Fig.~\ref{fig:member-5D}, the histogram of parallax for these cluster members is fitted with a Gaussian distribution, $\mu=0.86$ mas, $\sigma=0.08$ mas.
The typical uncertainties in parallax are $0.02-0.1$ mas for stars brighter than $G=18$ mag \citep{GaiaCollaboration+etal+2018}.
The parallaxes of these variable members are shown as blue dots, and they are less than $2\sigma$ of the Gaussian distribution.
Values of these parallaxes are in good agreement with that of NGC 1912 provided by \citet{Monteiro+Dias+2019} ($0.874 \pm 0.064$ mas).
 
 \begin{figure}[hp]
\begin{center} 
\includegraphics[height=5.cm,width=10.cm,clip,angle=0]{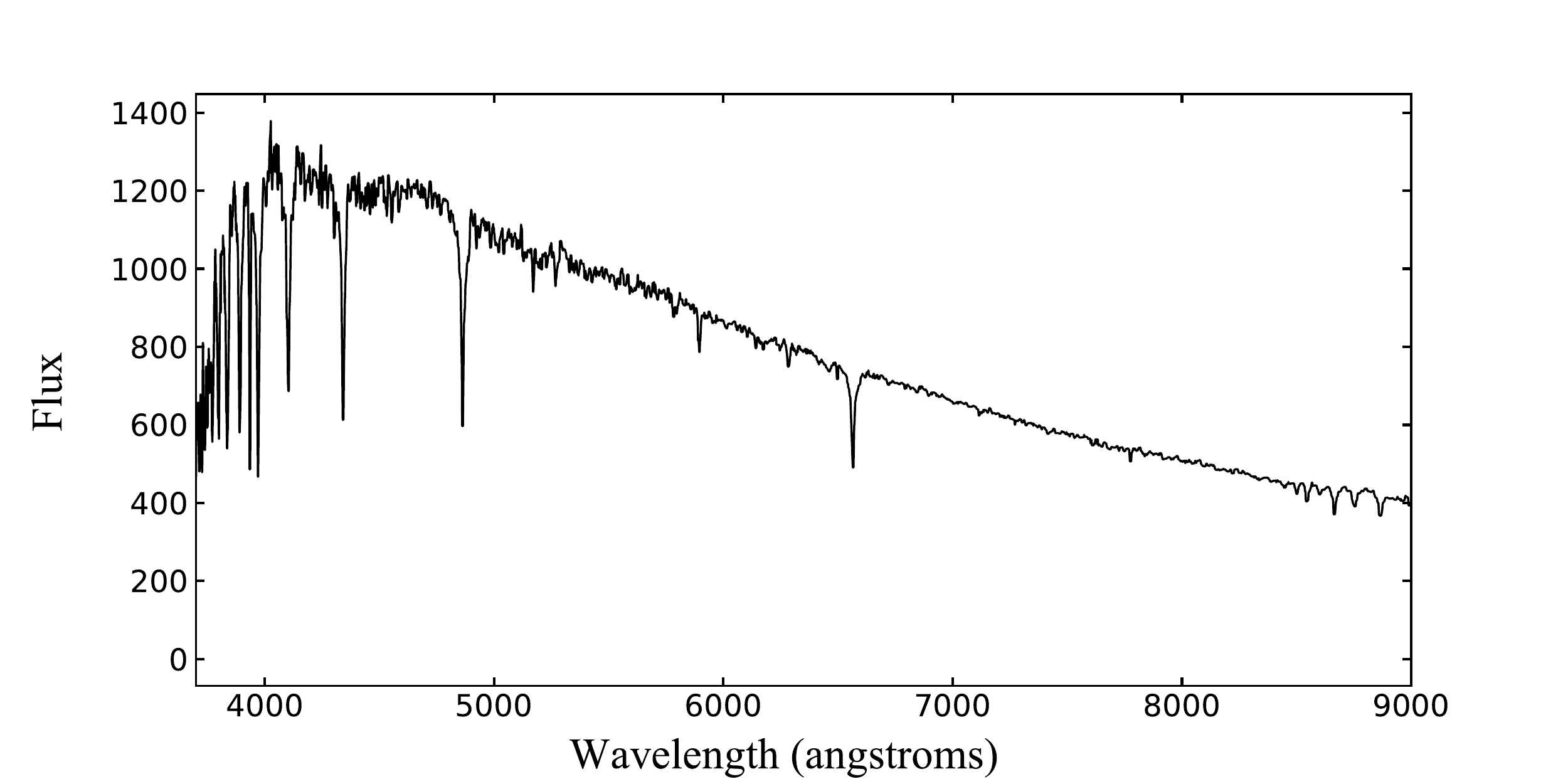}
\caption{Spectrum of V2.}
\label{fig:spec-V2}
\end{center}
\end{figure}

Among these variable cluster members, V2 is the only one that was observed by the Large Sky Area Multi-Object Fibre Spectroscopic Telescope (LAMOST) \citep{Cui+etal+2012, Zhao+etal+2012}.
Figure~\ref{fig:spec-V2} shows the LAMOST DR5 \citep{Luo+etal+2015} spectrum of V2. 
Note that the relative flux calibration was performed for the spectrum \citep{2012RAA....12..453S}, and there are no units for flux values.
The stellar parameters are $T_{eff}=7190\pm120$ K, $log(g)=4.2\pm 0.2$, $[Fe/H]=-0.3\pm 0.1$ and heliocentric radial velocity $R_{v}=-5\pm10$ km/s.
The parameters of V2 are in good agreement with that of $\gamma$\,Dor stars in literature.
Considering that V1...V5 and V14 were identified as cluster members with membership probabilities large than 50\% by \citet{Cantat-Gaudin+etal+2018}, the parameters of the other variable members should be very likely similar to that of V2.

These 24 detected variable stars are listed in the input catalog of TESS, but the observation data of these variable stars have not been released.

\subsection{Another probable variable cluster member }\label{sec:other}
Among the 24 variable stars we detected, six variables were reported as the cluster members of NGC 1912 by \citet{Cantat-Gaudin+etal+2018}.
Apart from these six variable cluster members, there is one variable star that seems to be another variable cluster member, which is V10.
It is a variable star in the field of NGC 1912 and exhibits a similar motion with the cluster members whose membership probabilities are larger than 0.9.

As shown in the panel (a) and panel (c) of Fig.~\ref{fig:member-5D}, V10 is located within the field of the open cluster.
The parallax of V10 is $\omega_{V10}=1.0644\pm0.044$ mas , which is between 2 and 2.5 $\sigma$ of the cluster's parallax distribution.

From kinematic information, we may get more hints about the membership of V10.
Since there are no radial velocity measurements for the cluster members and detected variables in {\it Gaia} DR2, we can use proper motion parameters to perform the kinematic analysis only.
The cluster members and detected variables are all brighter than $G=18$ mag, corresponding to typical uncertainties 0.3 mas$\,\mathrm{yr^{-1}}$ in proper motion \citep{Cantat-Gaudin+etal+2018}.
As shown in the panel (b) of Fig.~\ref{fig:member-5D}, V10 shows a similar proper motion with the variable cluster members.
The panel (d) of Fig.~\ref{fig:member-5D} shows the spatial distributions and corresponding tangential velocities of the detected variable stars and the cluster members with membership probabilities over 90\%.
The distances of these stars were estimated by \citet{Bailer-Jones+etal+2018}.
Thus, we can calculate their tangential velocities ($V_{t}$) from their proper motions and distances:
\begin{equation}
V_{t}=4.74 \times \mu \times d
\label{eq:vt}
\end{equation}

where $\mu$ is the proper motion in arcsec/year, d is the distance in parsecs, $V_{t}$ is the tangential velocity in km/s.
For the cluster members, their tangential velocities range from 6.82 km/s to 12.05 km/s in RA, and -28.16 km/s to -24.24 km/s in Dec.
As demonstrated in the panel (d) of Fig.~\ref{fig:member-5D}, the magnitude and direction of the tangential velocity of V10 are almost the same as that of the cluster members whose membership probabilities are larger than 90\%.

It is possible that V10 is the another cluster member.
However, more observations are needed to determine the membership of V10.

\section{Summary} \label{sec: SM}
We investigated and characterized the variable stars in the open cluster NGC 1912 and its surrounding field by photometric observations.
We detected 24 variable stars from nine nights photometric observations.
We also referred the cluster members established by \citet{Cantat-Gaudin+etal+2018} and {\it Gaia} DR2 to obtain the membership probability and the physical parameters of these detected variable stars.
Six variable cluster members were found.
A new CMD was constructed based on the cluster members with membership probabilities over 50\%.

We classified the detected variable stars according to their main periods and the positions of the variable cluster members on the CMD.
We discovered six $\gamma$\,Dor stars, one $\delta$\,Sct star, one EW type star, and three EA type stars.
And we also confirmed 13 variable stars detected by \cite{Szabo+etal+2006} and \cite{Jeon+2009}, consisting of two $\gamma$\,Dor stars, seven $\delta$\,Sct stars, three EW type stars, and one unknown type variable star.
For the unknown type variable star, its spectrogram and light curves show that it may be a long-term variable star.
More observations are needed to determine the period of it.
In the future, the more TESS data released will help us further understand the variable stars in the open cluster NGC 1912.

In addition to the six variable members reported by \citet{Cantat-Gaudin+etal+2018}, we potentially find a new cluster member candidate.
It can be confirmed by the homogeneity of the position and kinematics of the star with the identified cluster members.
We determined the main physical parameters of these variable cluster members by a fitting of isochrone.
The best fitting shows $log(age/yr)=8.75$, $[Fe/H]=-0.1$, $m-M=10.03$ mag, and $E($B$-$V$)= 0.307$.
These parameters are very similar to that of NGC 1912 provided by \cite{Monteiro+Dias+2019}.

\normalem
\begin{acknowledgements}
The authors acknowledge the National Natural Science Foundation of China under grants 11873081 and 11661161016, 2017 Heaven Lake Hundred-Talent Program of Xinjiang Uygur Autonomous Region of China, and the program of Tianshan Youth (No. 2017Q091).
The CCD photometric data of NGC 1912 were obtained with the Nanshan 1 m telescope of Xinjiang Astronomical Observatory.
This work has made use of data from the European Space Agency (ESA) mission {\it Gaia} (\url{https://www.cosmos.esa.int/web/gaia}), processed by the {\it Gaia} Data Processing and Analysis Consortium (DPAC, \url{https://www.cosmos.esa.int/web/gaia/dpac/consortium}).
Funding for the DPAC has been provided by national institutions, in particular the institutions participating in the {\it Gaia} Multilateral Agreement.
Guoshoujing Telescope (the Large Sky Area Multi-Object Fiber Spectroscopic Telescope LAMOST) is a National Major Scientific Project built by the Chinese Academy of Sciences.
Funding for the project has been provided by the National Development and Reform Commission. LAMOST is operated and managed by the National Astronomical Observatories, Chinese Academy of Sciences.

\end{acknowledgements}

\appendix 
\section{The results of frequency analysis for period variable stars}

\clearpage\footnotesize
\begin{longtable}{cccccc}
\caption{Frequency solutions for periodic variable stars.}\\
\label{tab:fa-V1-V24}\\
\hline
\hline
star ID &No. & $F_{19}$ & $A_{19}$ & $S/N_{19}$ & $F_{09}$\\
		 & &(c/d) &(mag) & & (c/d)\\
\hline
\endfirsthead
\hline
star ID &No. & $F_{19}$ & $A_{19}$ & $S/N_{19}$ & $F_{09}$\\
		 & &(c/d) &(mag) & & (c/d)\\
\hline
\endhead

\hline
\endfoot
V1&$f_{1}$ &$1.8350\pm0.0007$ & $0.0313\pm0.0006$&16.2&1.823\\
&$f_{2}$&$1.6160\pm0.0022$& $0.0123\pm0.0007$ &7.7&1.937\\
&$f_{3}$&$1.5100\pm0.0026 $& $0.0056\pm0.0008$&4.8&1.654\\
V2&$f_{1}$&$1.6432\pm0.0006 $ &$0.0448\pm0.0007$&8.4&1.644\\
&$f_{2}$ &-- &-- &-- & 2.310\\
V3&$f_{1}$&$1.7428\pm0.0019$ & $0.0203\pm0.0005$&23.9&--\\
&$f_{2}$&$2.6259\pm0.0059$&$0.0065\pm0.0005$ &6.4&--\\
&$f_{3}$&$1.8922\pm0.0041 $&$0.0065\pm0.0005$&8.6&--\\
V4&$f_{1}$&$2.2130\pm0.0008$ &$0.0191\pm0.0004$&14&--\\
&$f_{2}$&$2.4386\pm0.0018$& $0.0082\pm0.0004$&6.5&--\\
V5&$f_{1}$&$2.0076\pm0.0031$ &$0.0154\pm0.0006$&8.2&--\\
&$f_{2}$&$1.8948\pm0.0048$&$0.0052\pm0.0006$ &4.4&--\\
V6&$f_{1}$&$1.6526\pm0.0005$ &$0.0331\pm0.0004$&40.1&--\\
&$f_{2}$&$3.3152\pm0.0027$&$0.0059\pm0.0004$ &11.5&--\\
V7&$f_{1}$&$3.1553\pm0.0003$ &$0.0324\pm0.0003$&36.1&--\\
&$f_{2}$&$0.06967\pm0.0018$ &$0.0041\pm0.0003$&5.3&--\\
&$f_{3}$&$4.7281\pm0.0036$ &$0.0030\pm0.0003$&4.8&--\\
V8&$f_{1}$&$4.9512\pm0.0032$ &$0.0056\pm0.0004$&7.4&--\\
&$f_{2}$&$0.3716\pm0.0051$&$0.0040\pm0.0004$&4.7&--\\
V9&$f_{1}$&$8.8854\pm0.0014$ &$0.0109\pm0.0005$&18.0&8.878 \\
&$f_{2}$&$0.8958\pm0.0042$ &$0.0054\pm0.0005$&8.6&-- \\
&$f_{3}$&$8.3712\pm0.0041$ &$0.0046\pm0.0005$&8.1&-- \\
&$f_{4}$&$0.3086\pm0.0047$ &$0.0036\pm0.0006$&7.9&-- \\
&$f_{5}$&$8.6299\pm0.0059$ &$0.0035\pm0.0004$&4.8&-- \\
V10&$f_{1}$&$9.2986\pm0.0015$&$0.0121\pm0.0003$&13.5&9.301\\
&$f_{2}$&$8.4972\pm0.0029$&$0.0069\pm0.0003$&8.1&9.702\\
&$f_{3}$&$10.6937\pm0.0027$&$0.0049\pm0.0003$&5.8&7.596\\
&$f_{4}$&$6.0452\pm0.0041$ &$0.0046\pm0.0003$&4.6&--\\
V11&$f_{1}$&$19.9574\pm0.0031$ &$0.0063\pm0.0048$&8.1&19.965\\
&$f_{2}$&$3.2151\pm0.0064$ &$0.0042\pm0.0049$&5.2&--\\
&$f_{3}$&$2.0417\pm0.0055$ &$0.0034\pm0.0057$&5.0&--\\
&$f_{4}$&$4.0578\pm0.0067$ &$0.0026\pm0.0052$&4.8&--\\
V12&$f_{1}$& $17.1571\pm 0.0014$ &$0.0060\pm0.0002$ &26.8 & 17.152\\
&$f_{2}$& $0.7970\pm 0.0045$ &$0.0018\pm0.0002$ &5.1 & --\\
&$f_{3}$& $15.0934\pm 0.0063$ &$0.0014\pm0.0002$ &4.7 & --\\
&$f_{4}$& $19.5135\pm 0.0088$ &$0.0010\pm0.0002$&5.1 & --\\
V13&$f_{1}$&11.9810$\pm0.0019$ &$0.0049\pm0.0002$&16.9&11.980\\
&$f_{2}$&$15.2229\pm0.0037$ &$0.0027\pm0.0002$&5.7&--\\
&$f_{3}$&$1.9792\pm0.0041$ &$0.0027\pm0.0002$&4.3&--\\
&$f_{4}$&$12.8618\pm0.0049$ &$0.0020\pm0.0002$&4.8&--\\
V14&$f_{1}$&$21.7225\pm0.0035$ &$0.0052\pm0.0004$&7.6&21.719\\
&$f_{2}$&$28.0563\pm0.0062$ &$0.0027\pm0.0004$&4.6&--\\
V15&$f_{1}$&$8.9683\pm0.0013$ &$0.0323\pm0.0010$&22.1&--\\
&$f_{2}$&$16.1318\pm0.0038$&$0.0114\pm0.0010$ &5.5&--\\
&$f_{3}$&$6.9378\pm0.0029$&$0.0108\pm0.0010$ &9.5&--\\
&$f_{4}$&$0.0630\pm0.0021$&$0.00980.0010$ &16.4&--\\
&$f_{5}$&$0.8029\pm0.0037$&$0.00790.0010$ &9.3&--\\
&$f_{6}$&$8.6233\pm0.0050$&$0.00720.0010$ &5.4&--\\
V23&$f_{1}$&$1.7732\pm0.0062$ &$0.0253\pm0.0006$&47.8&--\\
&$f_{2}$&$2.0048\pm0.0049$ &$0.0112\pm0.0006$&41.9&--\\
&$f_{3}$&$1.4685\pm0.0084$ &$0.0107\pm0.0011$&32.9&--\\
&$f_{4}$&$1.1273\pm0.0066$ &$0.0065\pm0.0008$&25.3&--\\
&$f_{5}$&$2.3582\pm0.0099$ &$0.0061\pm0.0008$&13.5&--\\
&$f_{6}$&$3.5769\pm0.0077$ &$0.0039\pm0.0004$&7.7&--\\
&$f_{7}$&$3.9039\pm0.0012$ &$0.0033\pm0.0005$&5.4&--\\
&$f_{8}$&$4.3081\pm0.0023$ &$0.0027\pm0.0005$&5.9&--\\
&$f_{9}$&$0.3290\pm0.0087$ &$0.0021\pm0.0017$&6.4&--\\ 
V24&$f_{1}$&$0.2587\pm0.0035$ &$0.0057\pm0.0005$&5.9&--\\
&$f_{2}$&$2.5747\pm0.0048$ &$0.0037\pm0.0005$&5.2&--\\ 
\hline
\end{longtable}
\tablecomments{0.86\textwidth}{For each star, only independent frequencies are listed. The uncertainty of frequencies were calculated by the error matrix of the least-squares using Period04. A represents amplitude of frequency peaks in Fourier spectrum. The subscript of 09 and 19 mean that the Fourier analysis result of these variable stars derived by \cite{Jeon+2009} and this work, respectively.}

\section{The amplitude spectra of period variable stars}

 \begin{figure}
   \centering
   \includegraphics[height=8cm,width=7cm,clip,angle=0]{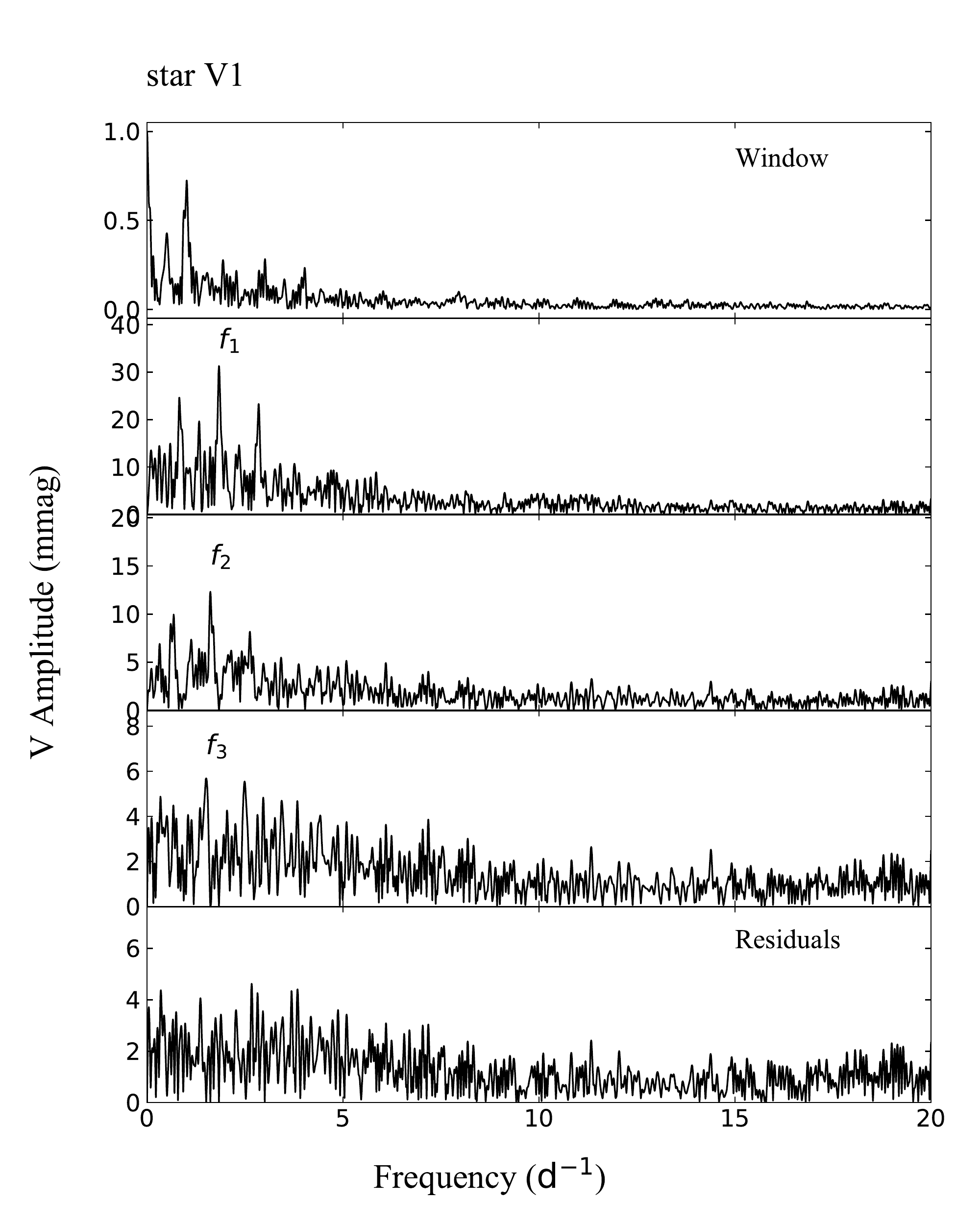}
      \includegraphics[height=8cm,width=7cm,clip,angle=0]{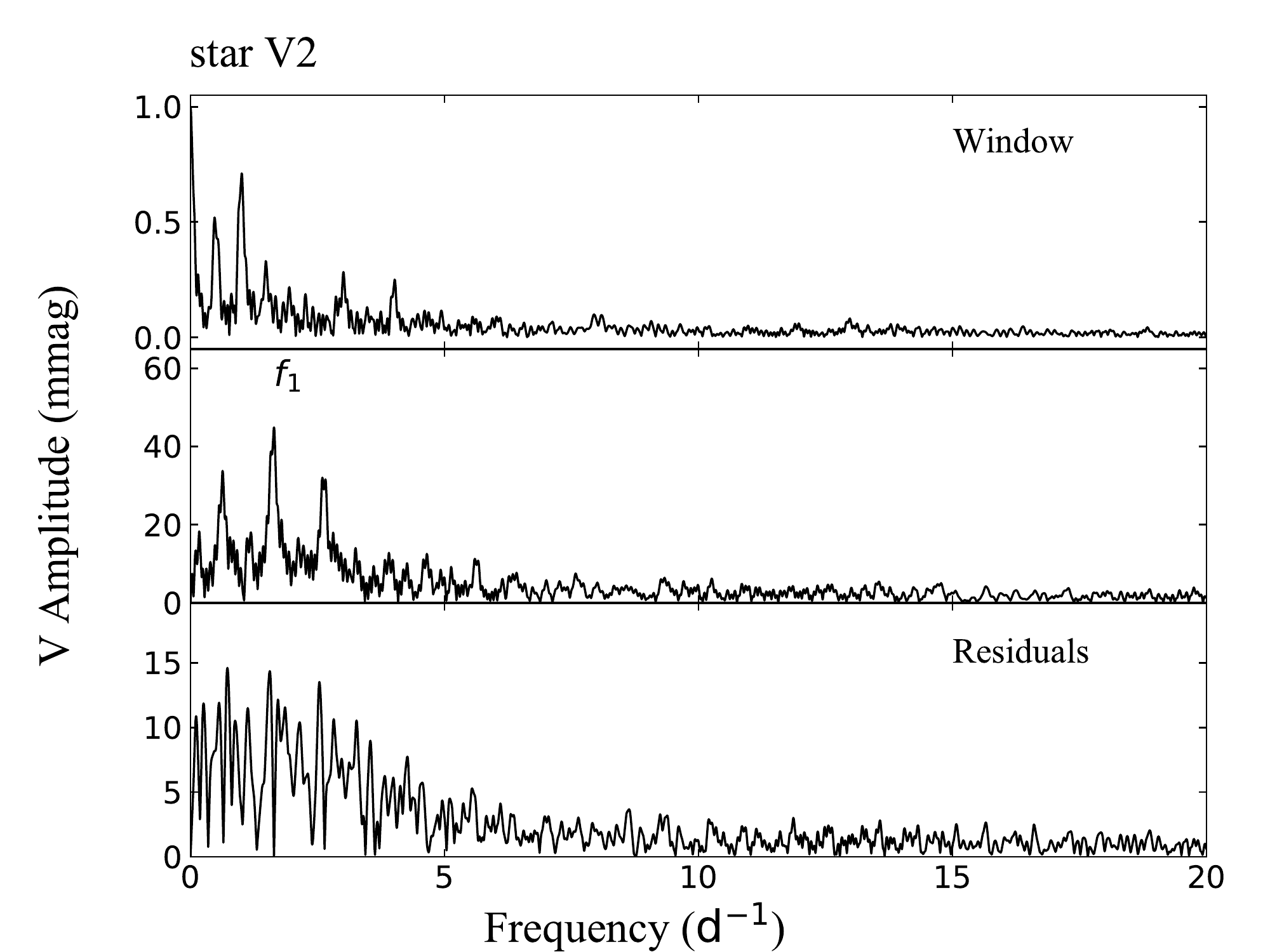}
         \includegraphics[height=8cm,width=7cm,clip,angle=0]{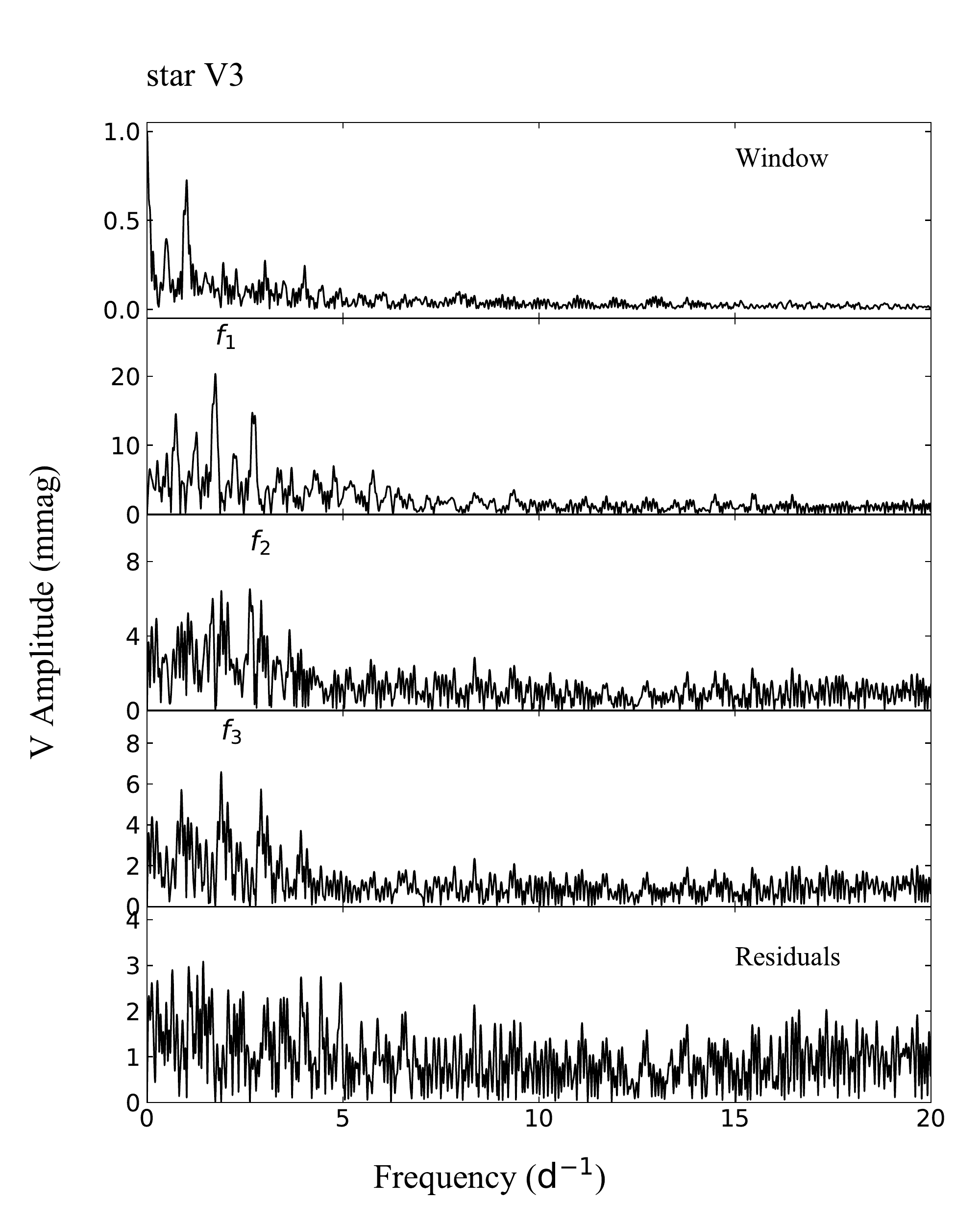}
            \includegraphics[height=8cm,width=7cm,clip,angle=0]{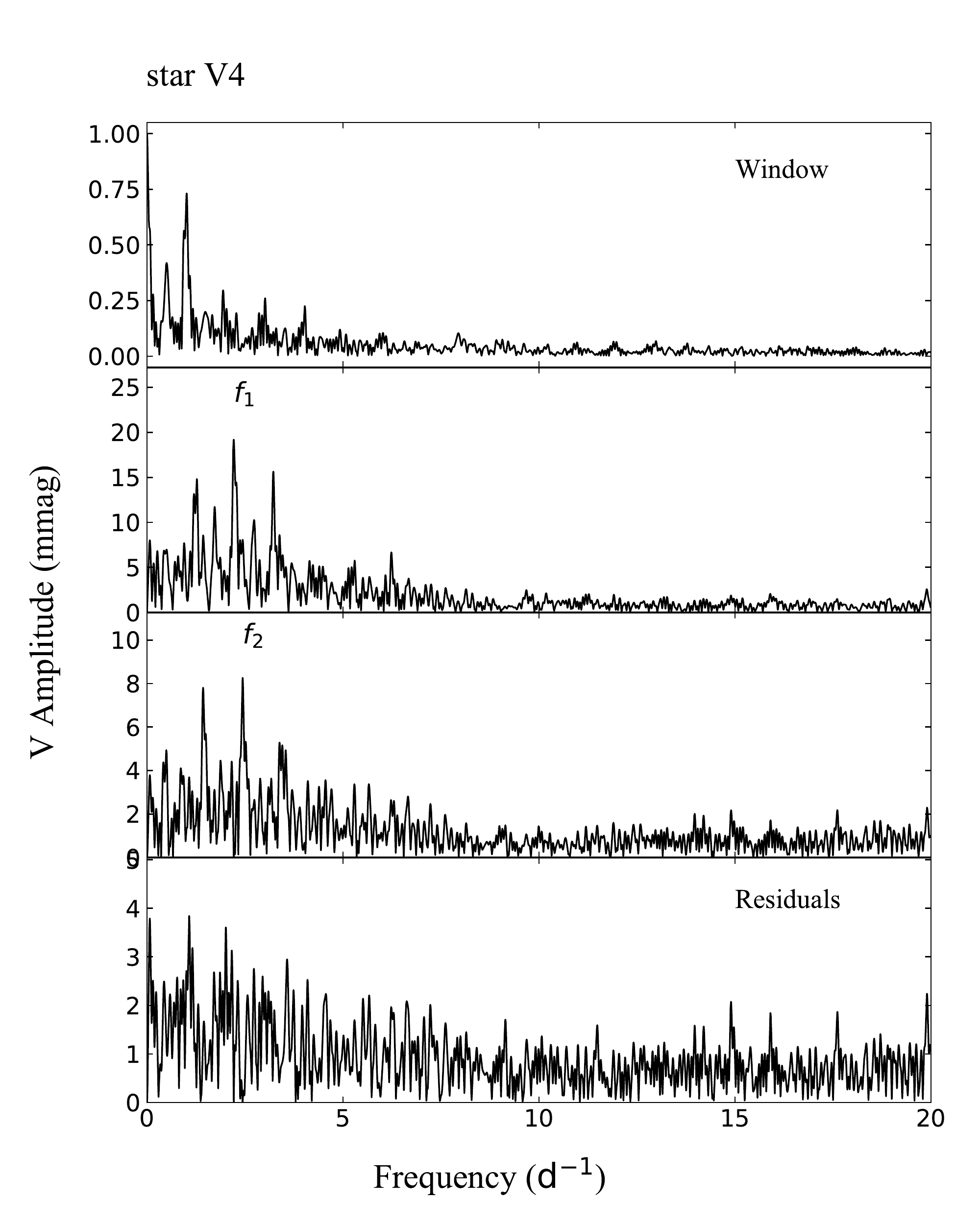}
   \caption{The same as Fig.~\ref{fig:as-0915} but for star V1 to V4.}
   \label{fig:as-V1-V4}
   \end{figure}
   
    \begin{figure}
   \centering
   \includegraphics[height=8cm,width=7cm,clip,angle=0]{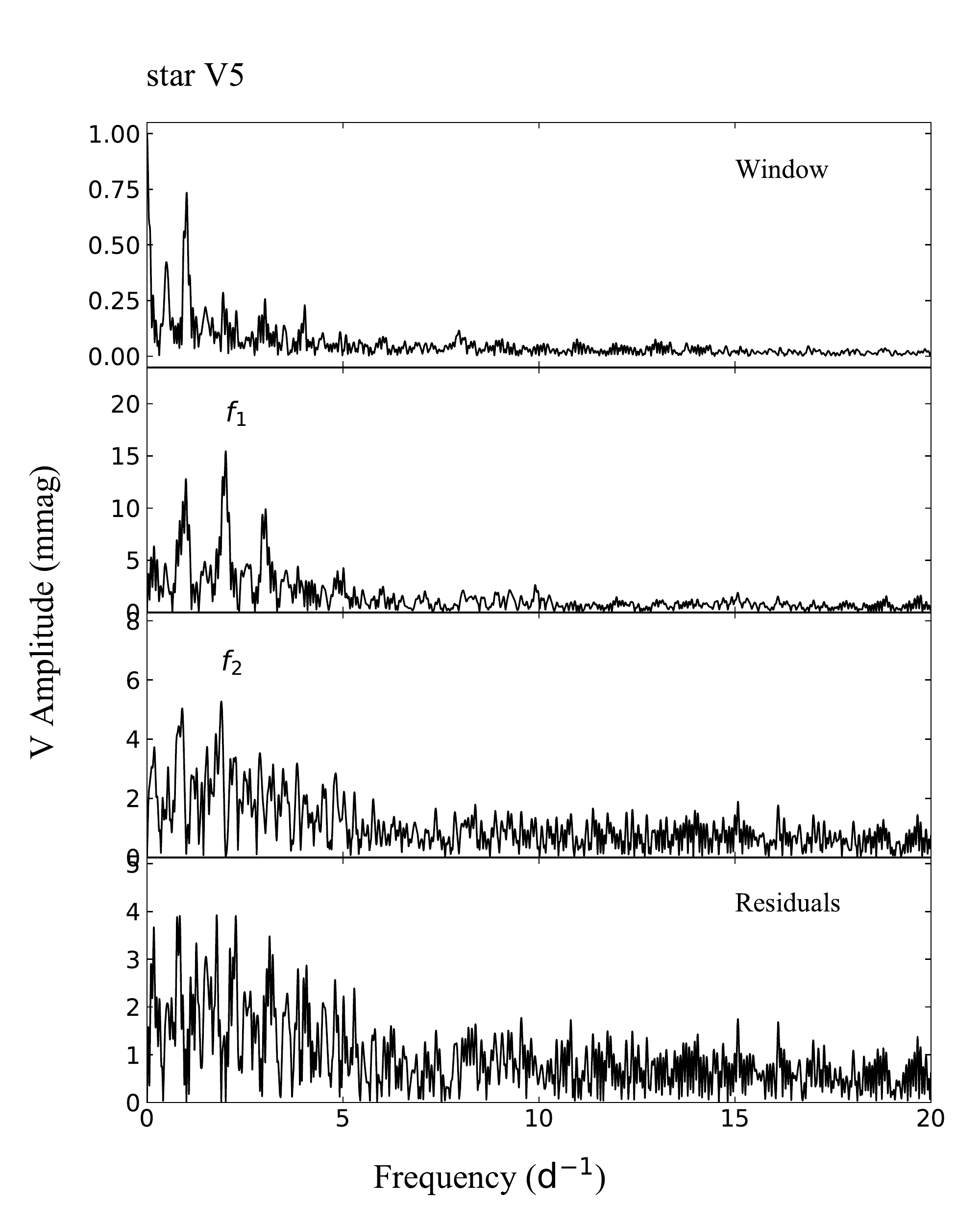}
      \includegraphics[height=8cm,width=7cm,clip,angle=0]{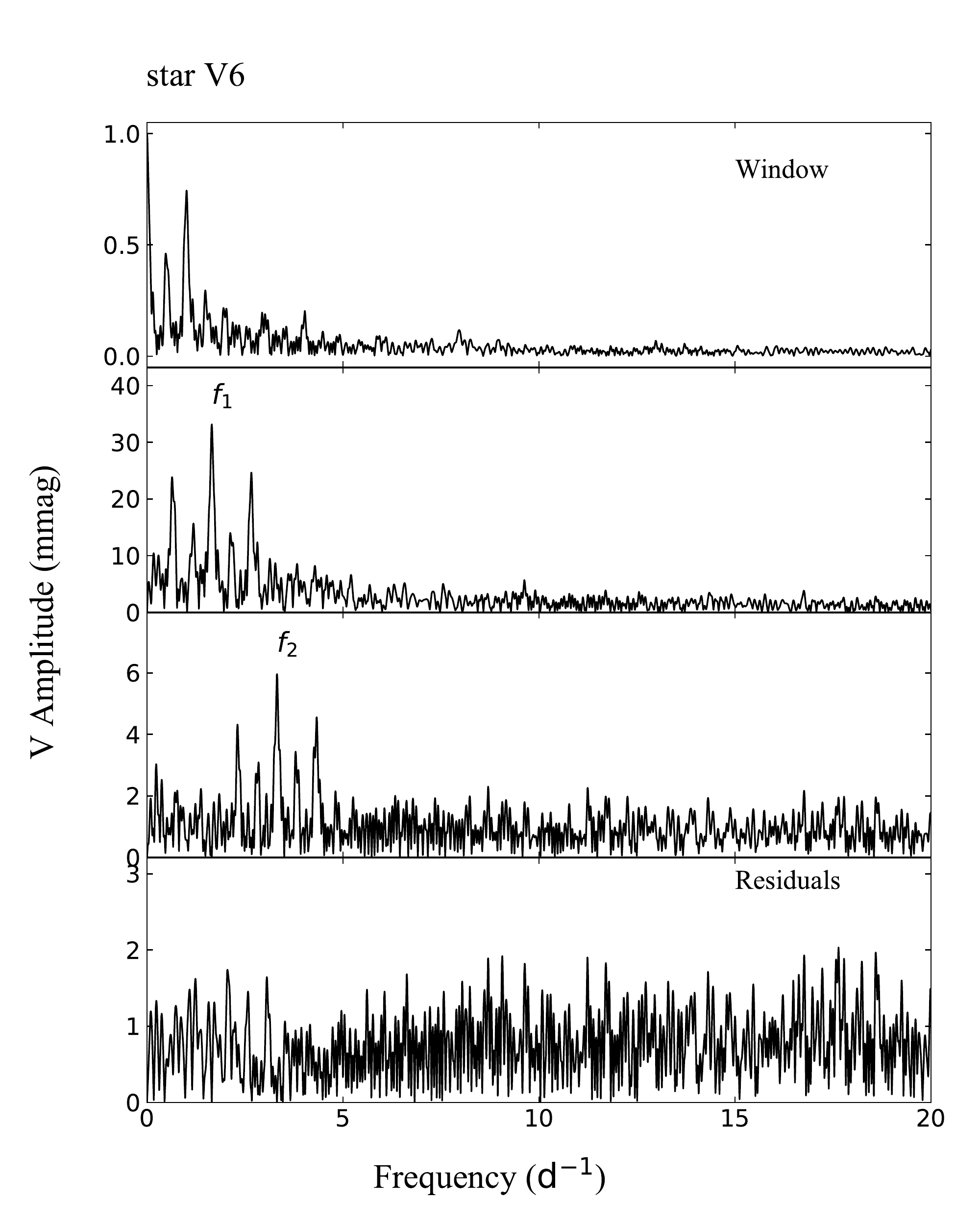}
         \includegraphics[height=8cm,width=7cm,clip,angle=0]{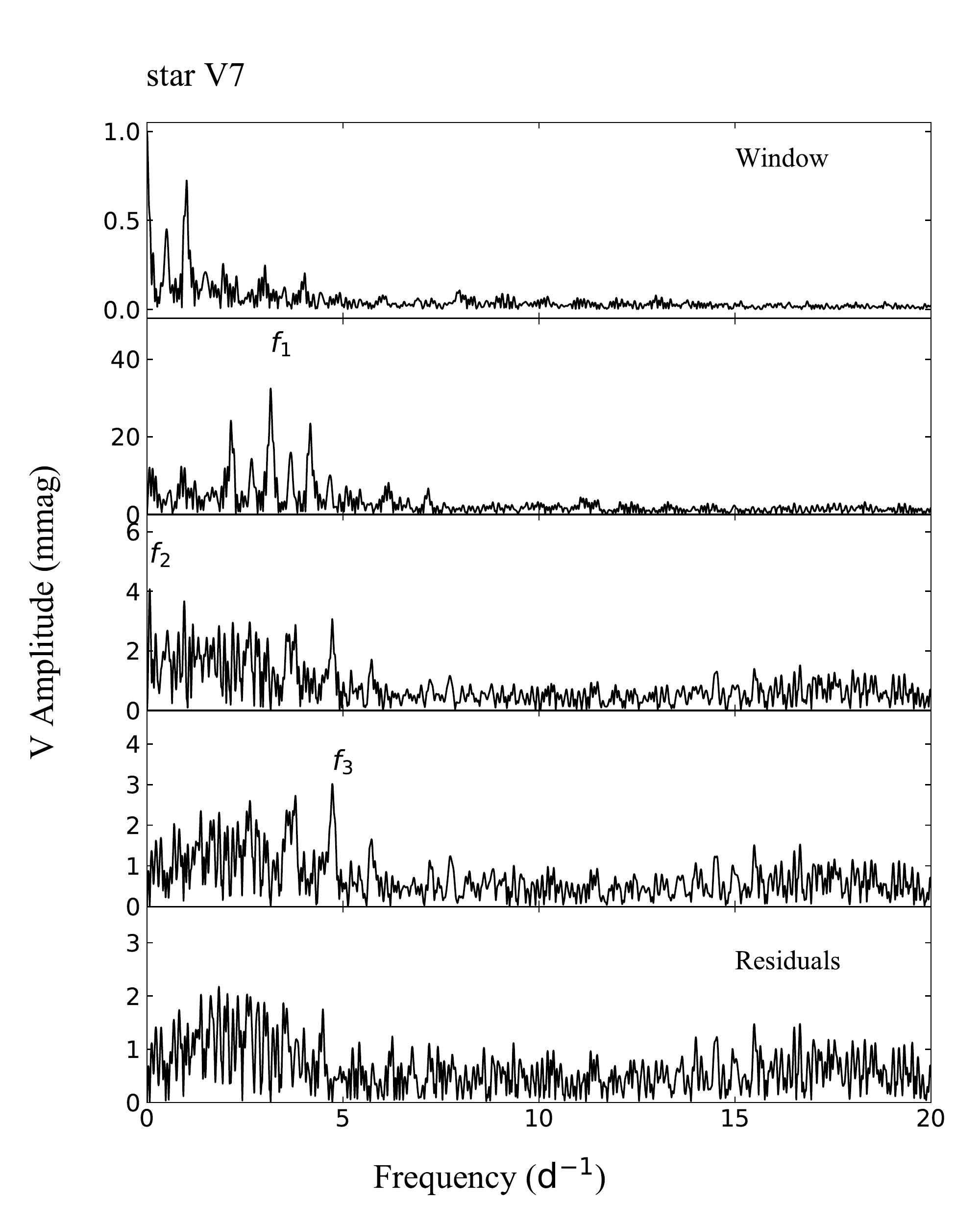}
   \caption{The same as Fig.~\ref{fig:as-0915} but for star V5 to V7.}
   \label{fig:as-V5-V7}
   \end{figure}
   
   \begin{figure}
   \centering
   \includegraphics[height=10cm,width=7cm,clip,angle=0]{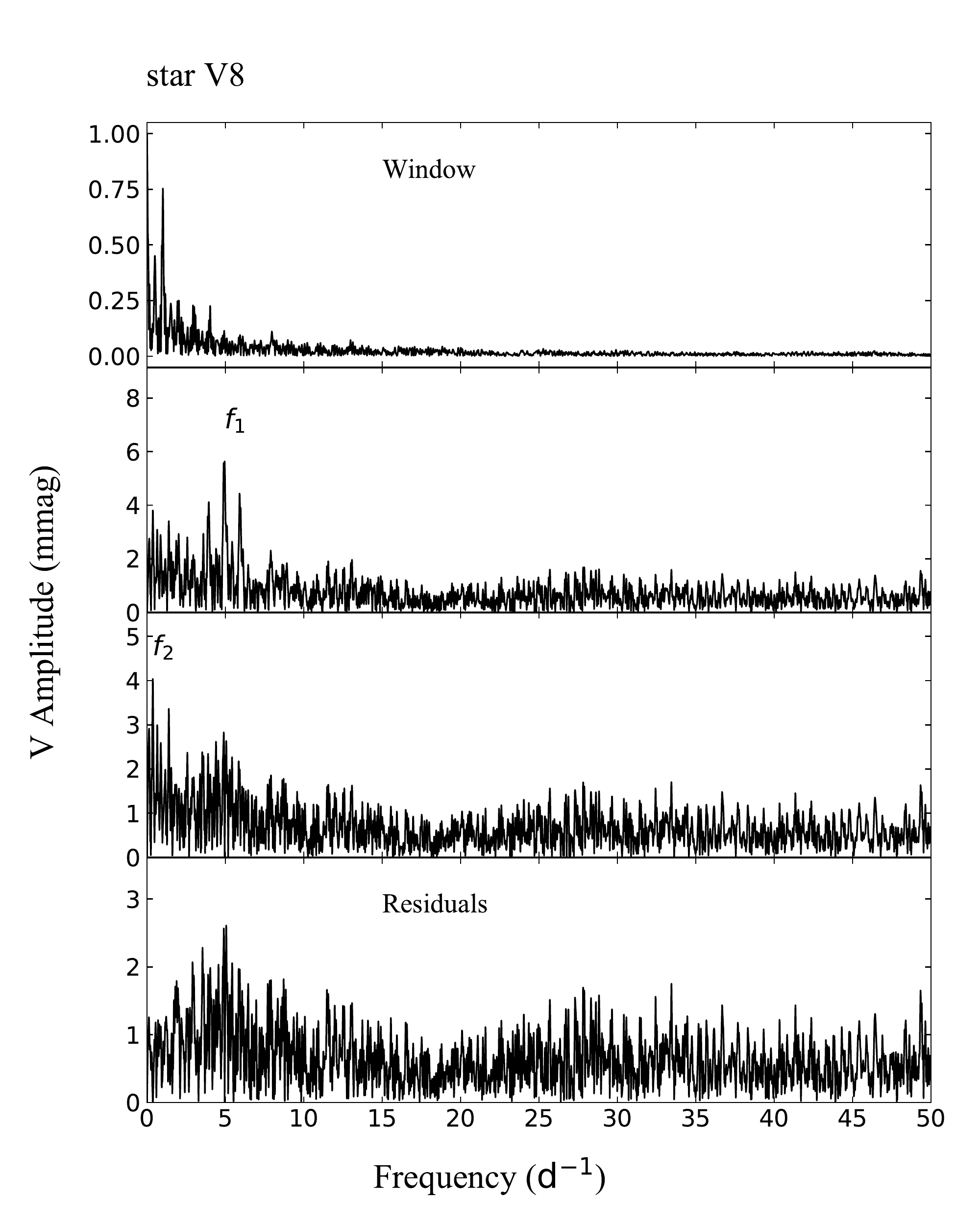}
   \includegraphics[height=10cm,width=7cm,clip,angle=0]{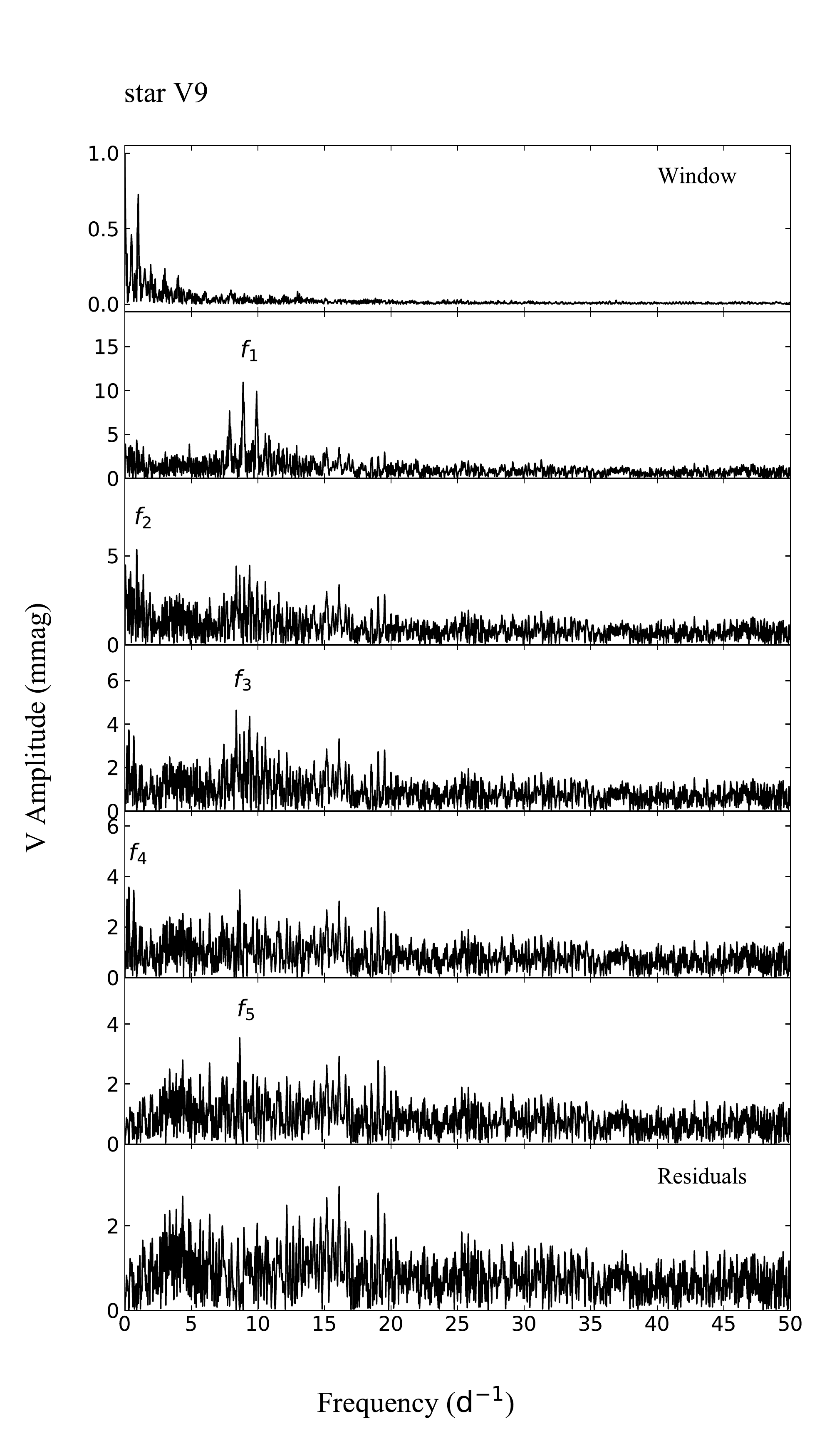}
      \includegraphics[height=10cm,width=7cm,clip,angle=0]{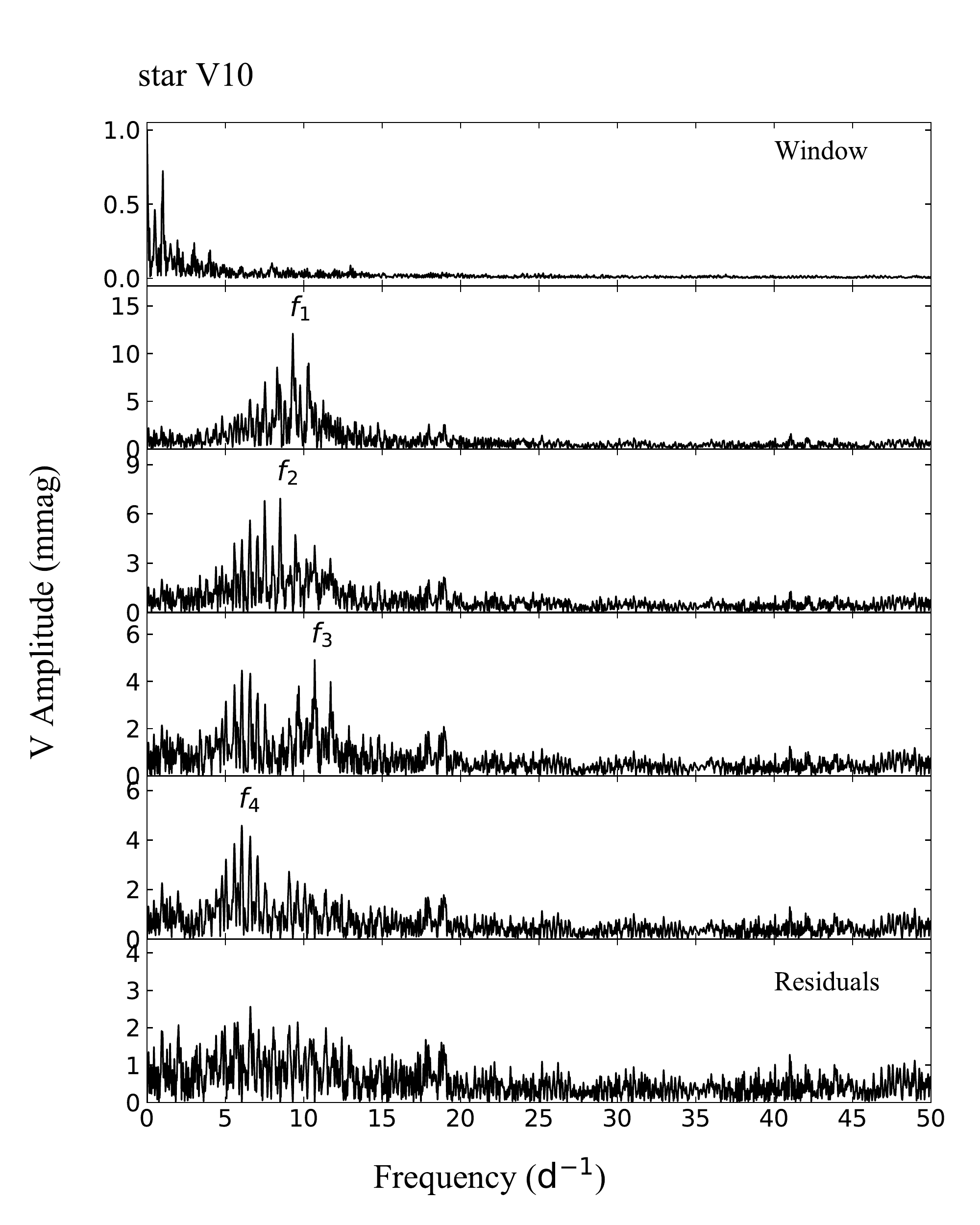}
         \includegraphics[height=10cm,width=7cm,clip,angle=0]{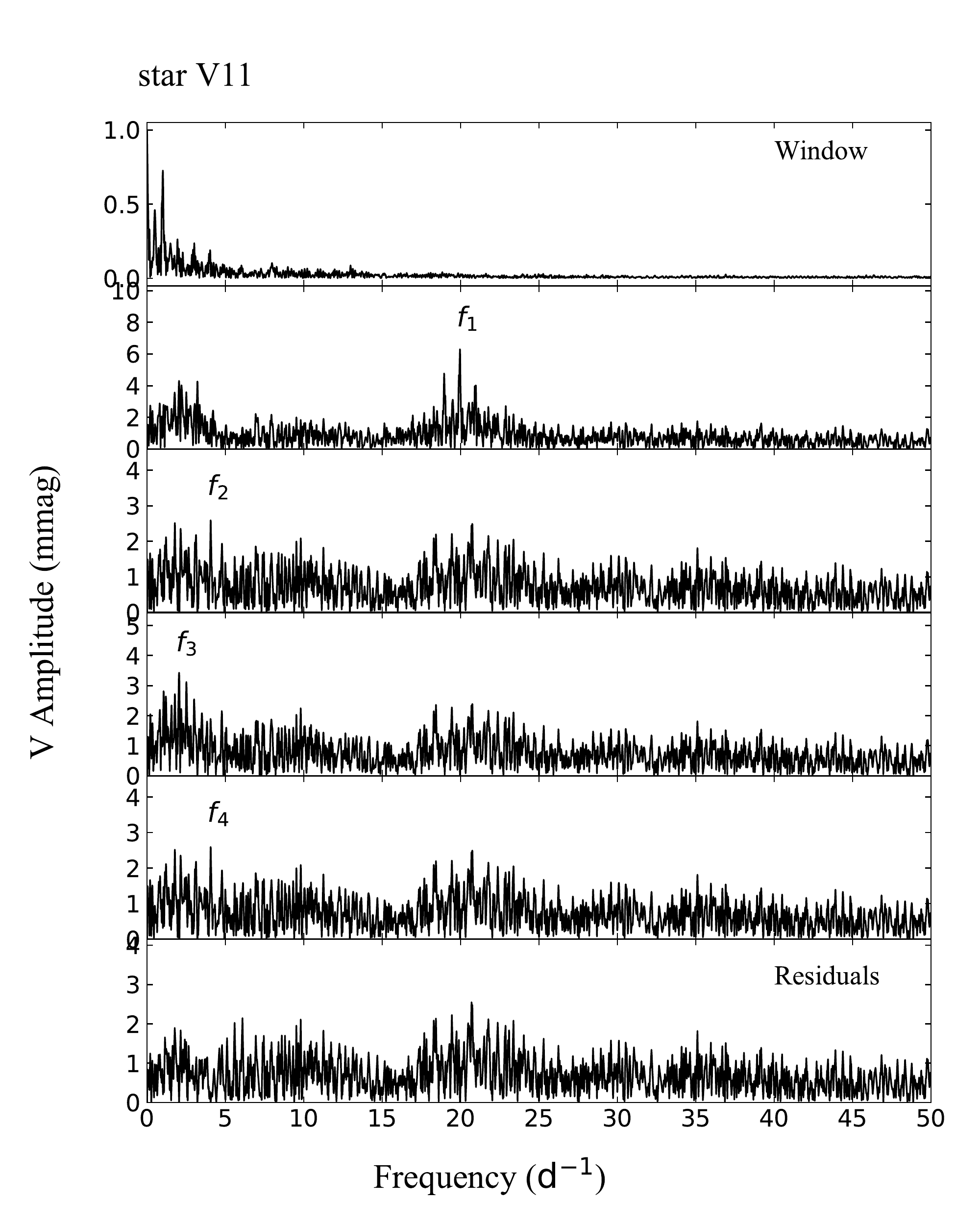}
   \caption{The same as Fig.~\ref{fig:as-0915} but for star V8 to V11.}
   \label{fig:as-V8-V11}
   \end{figure}
   
  \begin{figure}
   \centering
      \includegraphics[height=10cm,width=7cm,clip,angle=0]{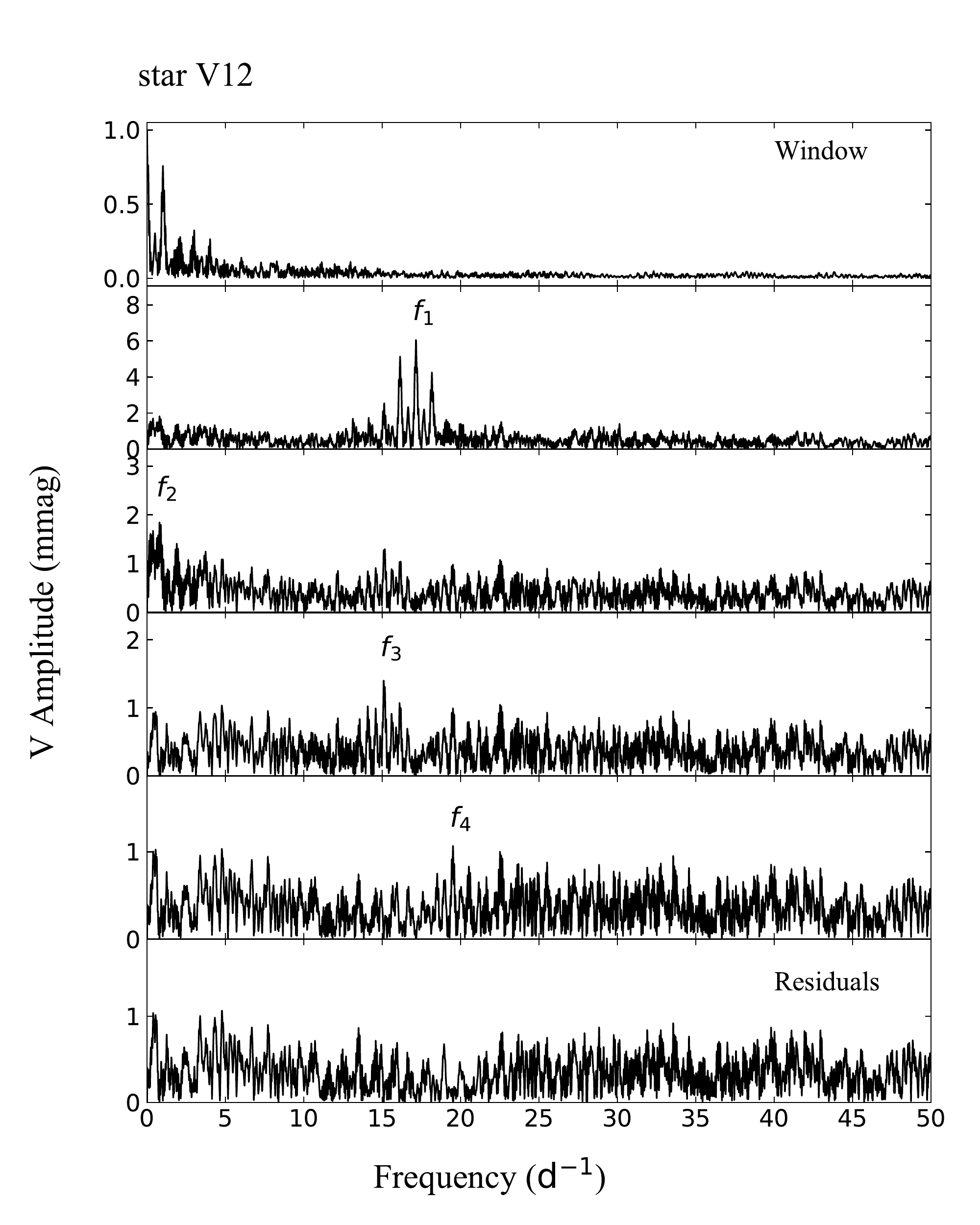}
      \includegraphics[height=10cm,width=7cm,clip,angle=0]{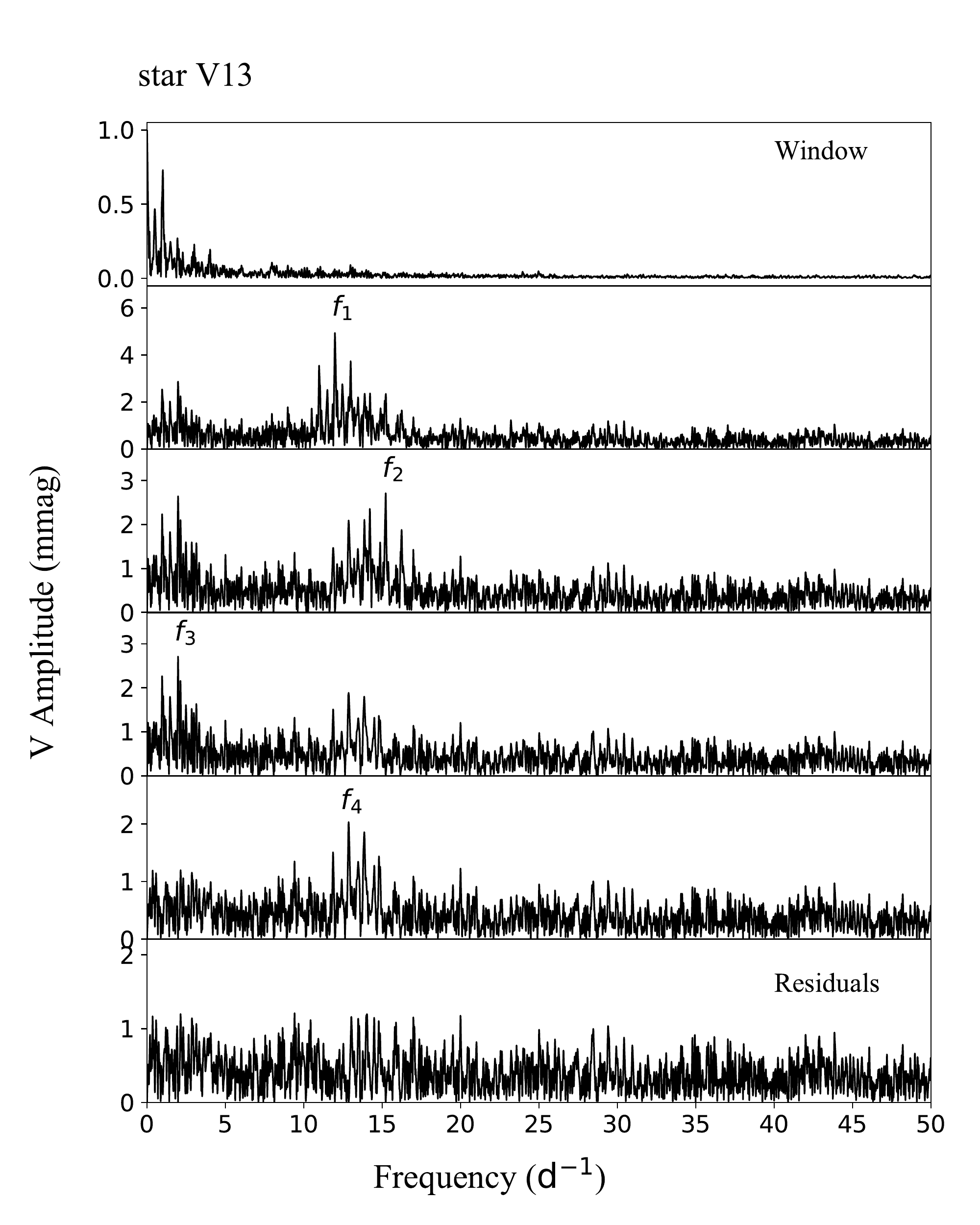}
   \caption{The same as Fig.~\ref{fig:as-0915} but for star V12 and V13.}
   \label{fig:as-V12-V14}
   \end{figure}
   
     \begin{figure}
   \centering
      \includegraphics[height=10cm,width=7cm,clip,angle=0]{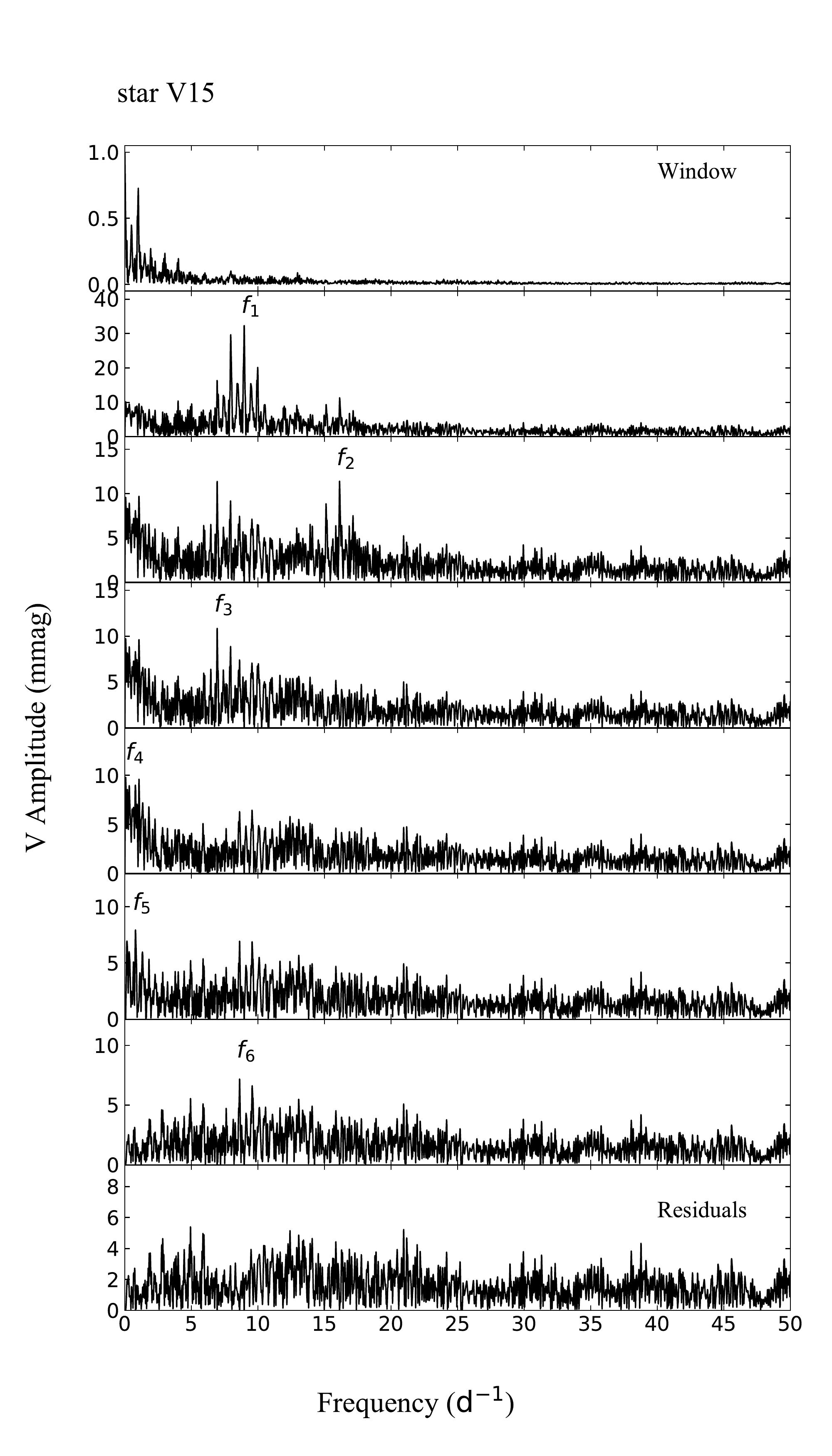}
   \caption{The same as Fig.~\ref{fig:as-0915} but for star V15.}
   \label{fig:as-V15}
   \end{figure}
      
   \begin{figure}
   \centering
   \includegraphics[width=1\linewidth, angle=0]{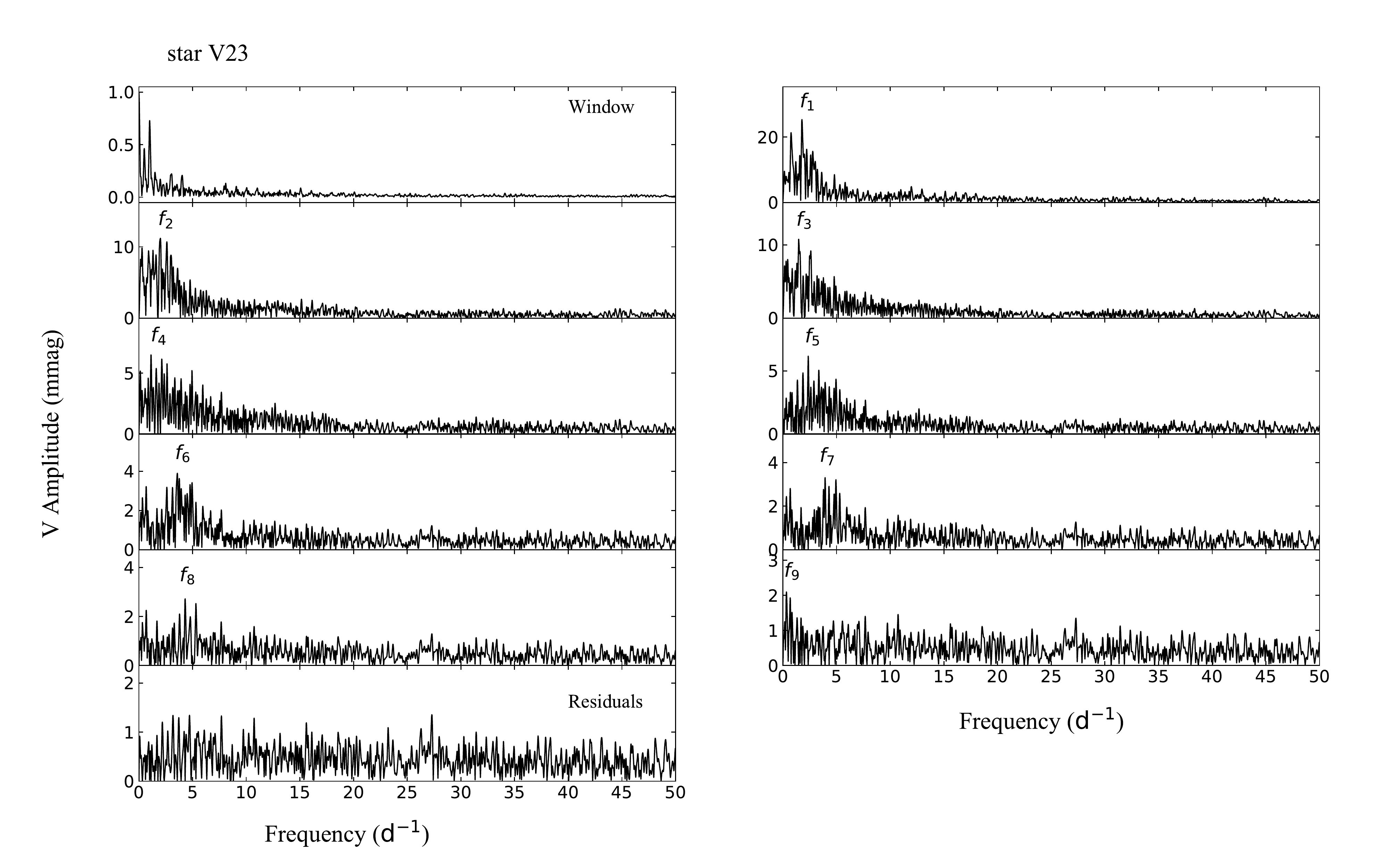}
   \caption{The same as Fig.~\ref{fig:as-0915} but for star V23.}
   \label{fig:as-V23}
   \end{figure}
   
  \begin{figure}[hp]
 \centering
 \includegraphics[height=10cm,width=9cm,clip,angle=0]{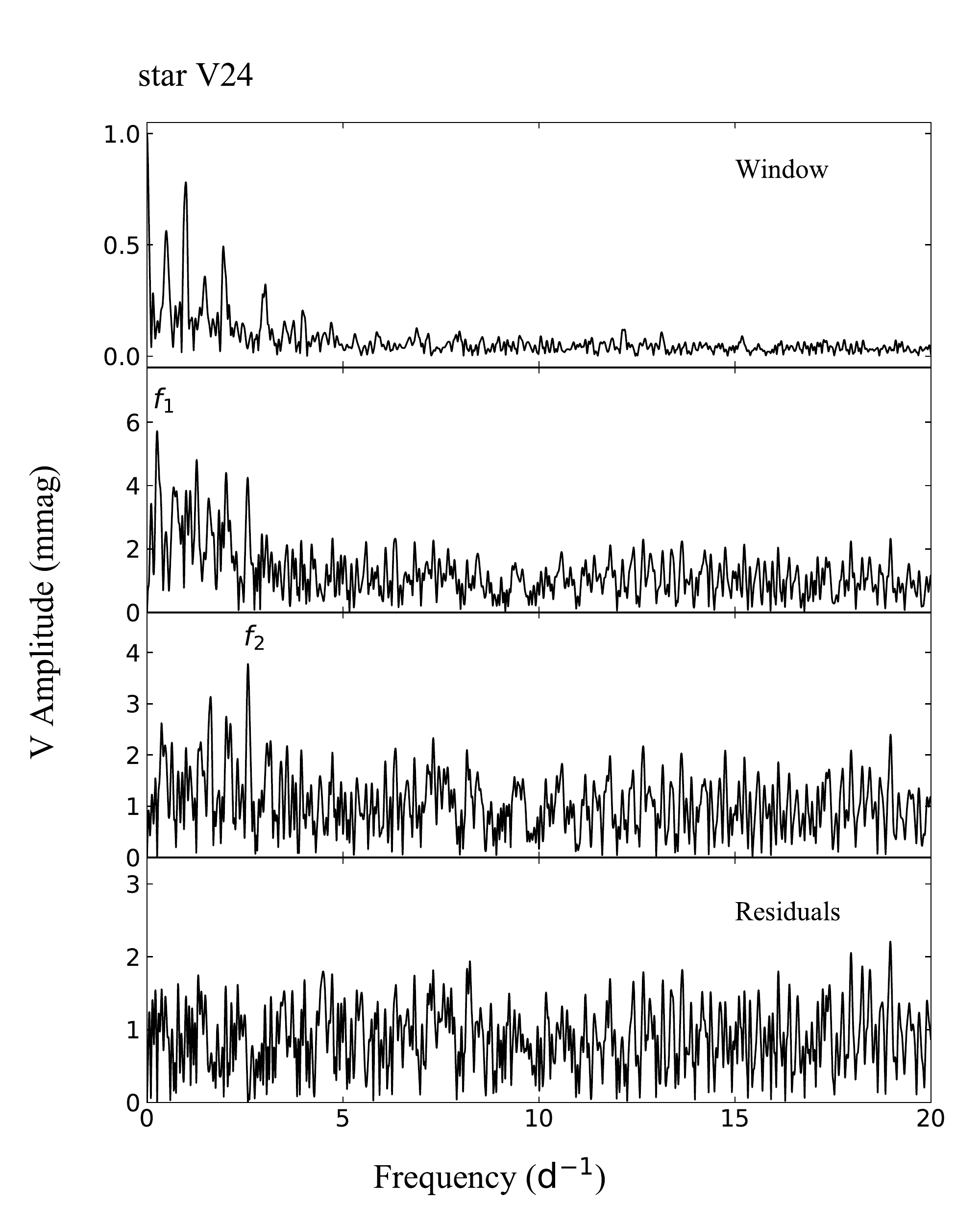}
 \caption{The same as Fig.~\ref{fig:as-0915} but for star V24.}
 \label{fig:as-V24}
\end{figure} 
  
\bibliographystyle{raa}
\bibliography{main}

\end{document}